\documentclass[fleqn,usenatbib,useAMS]{mnras}

\usepackage{amsmath}                
\usepackage{amsfonts}               
\usepackage{amssymb}                

\usepackage[T1]{fontenc}
\usepackage{aecompl}

\usepackage{longtable}
\usepackage{graphicx}

\providecommand{\tabularnewline}{\\}

\makeatother

\title[Circumstellar triple common envelope evolution]{Simulations of common envelope evolution in triple systems: Circumstellar case}

\author[Glanz \& Perets]{
	Hila Glanz$^{1}$ \thanks{E-mail: glanz@tx.technion.ac.il} and Hagai B. Perets $^{1}$
	\\
	$^{1}$Technion - Israel Institute of Technology, Haifa, 3200002, Israel}

\date{Accepted 2020 October 14. Received 2020 October 14; in original form 2020 March 31}

\pubyear{2020}

\begin{document}
\label{firstpage}
\pagerange{\pageref{firstpage}--\pageref{lastpage}}
\maketitle

\begin{abstract}
The dynamical evolution of triple stellar systems could induce the formation of compact binaries and binary mergers. Common envelope (CE) evolution, which plays a major role in the evolution of compact binary systems, can similarly play a key role in the evolution of triples. Here we use hydrodynamical simulations coupled with few-body dynamics to provide the first detailed models of triple common envelope (TCE) evolution. We focus on the circumstellar case, where the envelope of an evolved giant engulfs a compact binary orbiting the giant, which then in-spirals into the core of the evolved star. Through our exploratory modeling we find several possible outcomes of such TCE: (1)  The merger of the binary inside the third star's envelope; (2) The disruption of the in-spiraling binary following its plunge, leading to a chaotic triple dynamics of the stellar-core and the two components of the former disrupted binary. The chaotic evolution typically leads to the in-spiral and merger of at least one of the former binary components with the core, and sometimes to the ejection of the second, or alternatively its further now-binary common-envelope evolution. The in-spiral in TCE leads to overall slower in-spiral, larger mass ejection and the production of more aspherical remnant, compared with a corresponding binary case of similar masses, due to the energy/momentum extraction from the inner-binary. We expect TCE to play a key role in producing various types of stellar-mergers and unique compact binary systems, and potentially induce transient electromagnetic and gravitational-wave sources.
\end{abstract}


\section{Introduction}
Triple systems are frequent among stellar systems, and in particular massive systems {\citep[e.g.][and references therein]{toonen2016evolution,Moe+17}}, and their evolution may lead to a wide variety
of non-trivial and sometimes exotic outcomes. These include the formation
of various types of compact stellar binaries and triples,
stellar mergers, and the possible triggering of transient events (e.g.
\citealp{1986MNRAS.220P..13E,Iben+99,Ford+00,Soker04,2009ApJ...697.1048P,2011ApJ...741...82T,per+12c,Ham+13,Tauris_2014,2015MNRAS.450.1716S,2016ARA&A..54..441N,DiStefano2019}).
Although triple stellar systems had been extensively studied, the
vast majority of the studies focused on the dynamical evolution of
such systems; either through short-term dynamical evolution of unstable
systems, or the longer-term secular evolution of triples {\citep[e.g. ][and references therein]{valtonen_karttunen_2006,2016ARA&A..54..441N}}.
Few studies explored the implications of mass-loss and/or mass transfer
in stellar triples (\citealp{Iben+99,Soker04,per+12c,2013ApJ...766...64S,deV+14,2014ApJ...794..122M,Hillel2017,stefano2018mass, Por+19};
see \citealp{toonen2016evolution} for an overview), but a detailed
modeling of the triple common envelope (TCE) phase, and in particular
fully hydrodynamical simulations of this phase had not yet been explored,
to the best of our knowledge, and are the focus of our study. 

Binary common envelope (CE) results from an unstable Roche-lobe
overflow in a binary system, most typically following the evolution
of one of the binary components and the extension of its envelope
during the red giant (RG) phase. The binary components are thought
to in-spiral inside the (now shared) CE, leading to the shrinkage
of the orbit, on the expense of the outer envelope expansion and possible ejection.
CE evolution (CEE) is believed to be one of the most important steps
in the evolution of close binaries, providing an essential part in
the formation of compact binaries and stellar mergers \citep[][and references therein]{pac76,Izz+12,Iva+13,Sok17}.

\begin{figure}
\centering
\includegraphics[width=0.8\linewidth,clip]{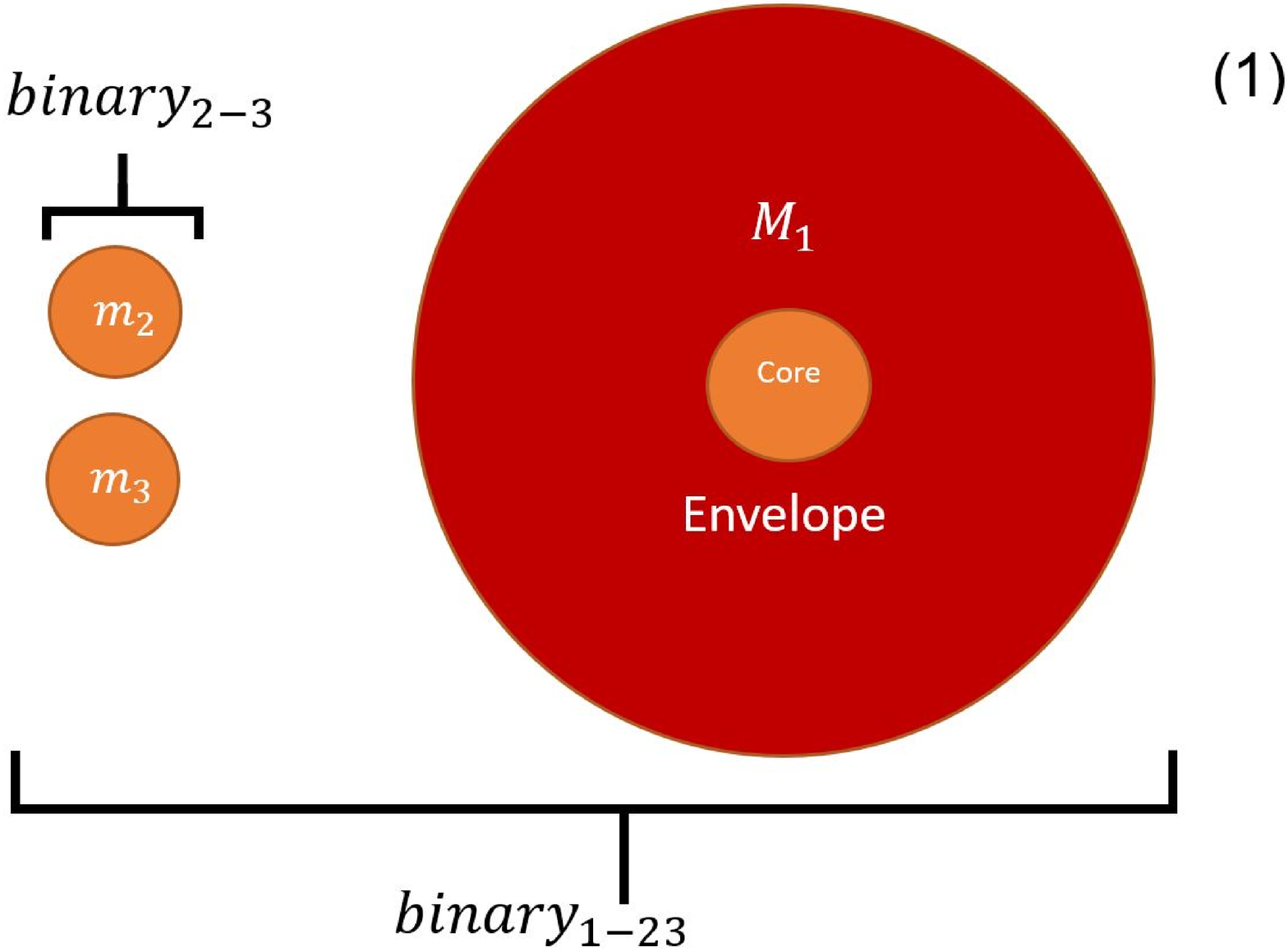}
\par
\includegraphics[width=0.8\linewidth,clip]{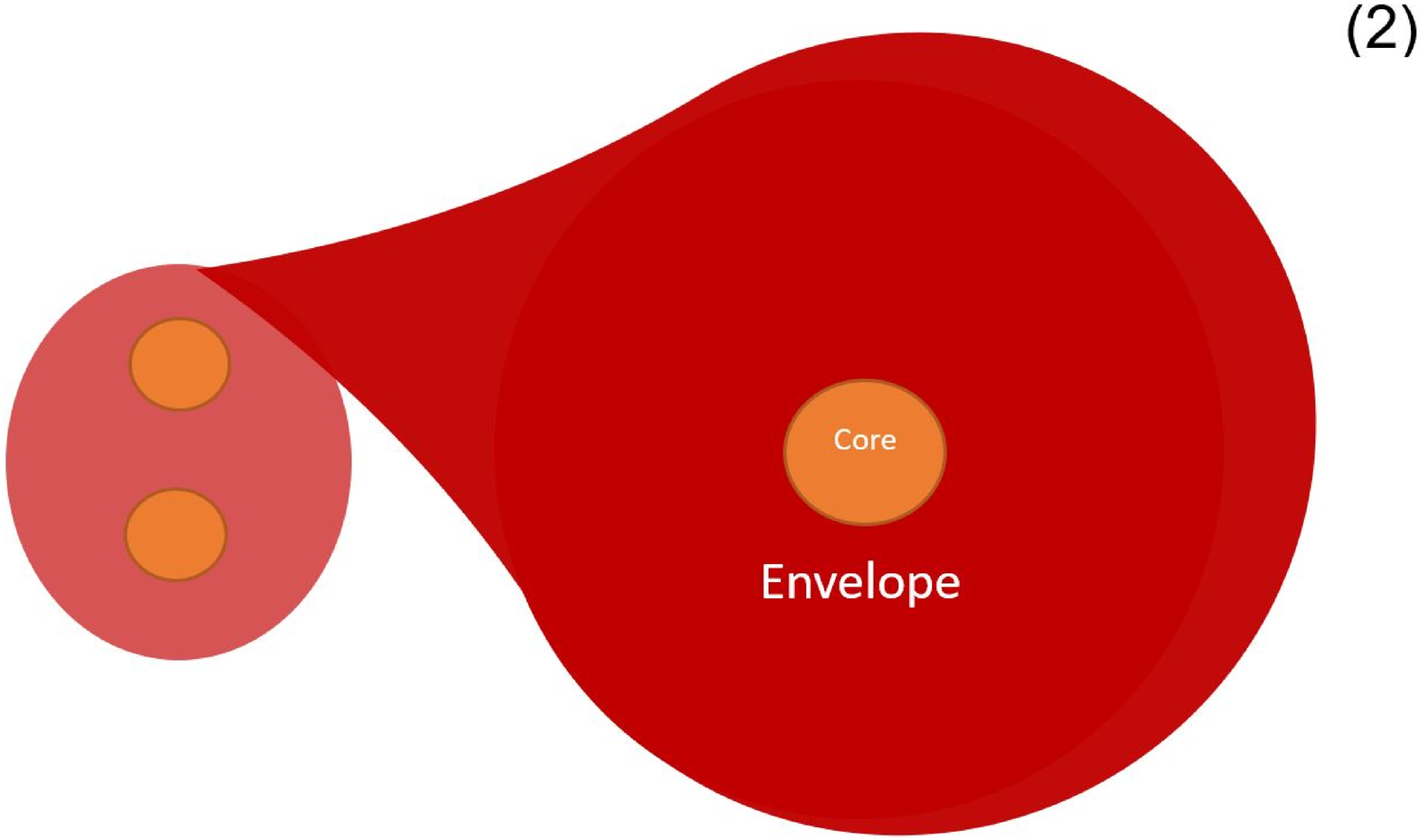}
\par
\includegraphics[width=0.8\linewidth,clip]{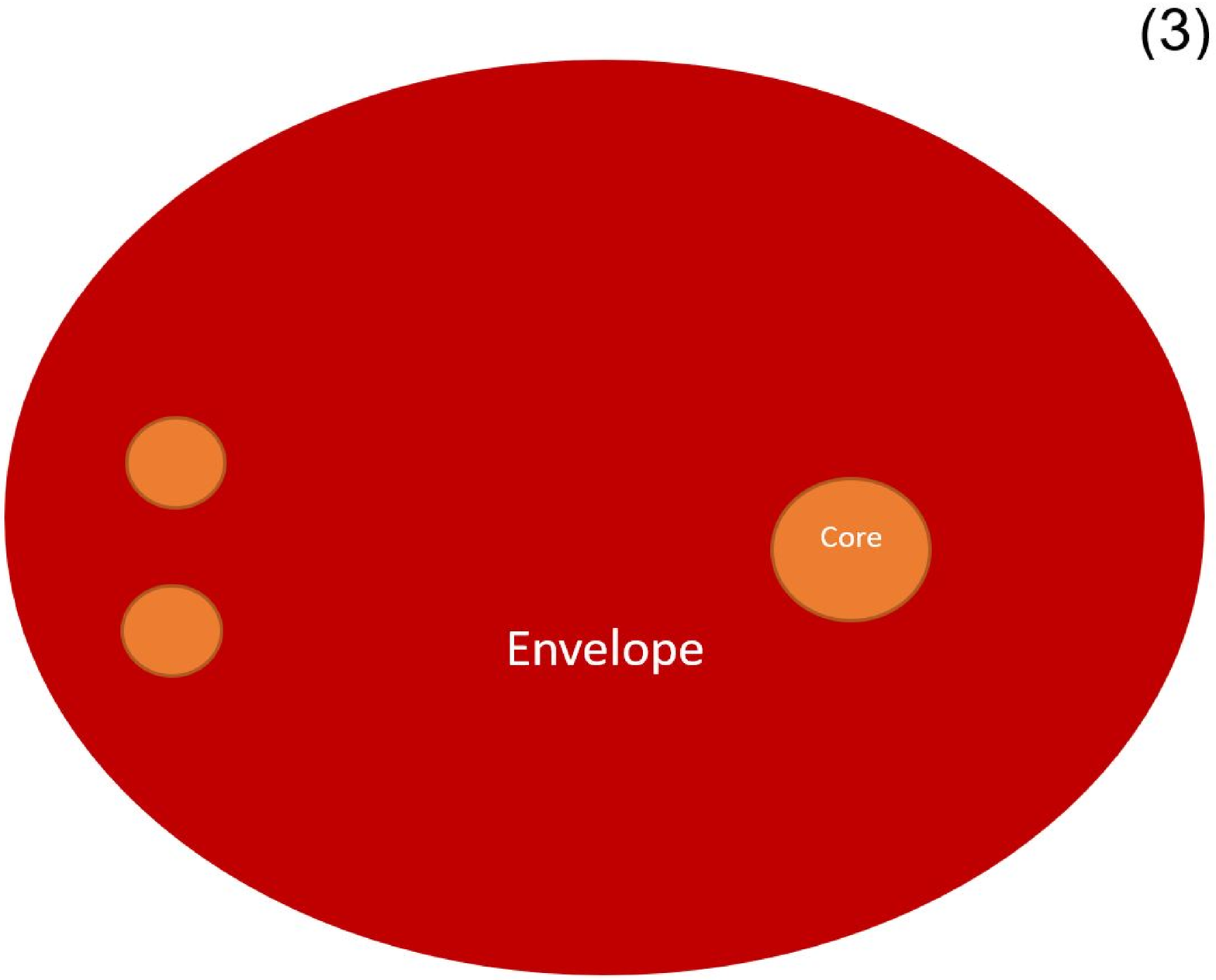}
\par
\caption{\label{fig:systemscatch}A schematic drawing of a circumstellar triple common envelope phase. The first stage introduces a hierarchical triple system prior to the Roche-lobe overflow, where a red-giant star is in orbit with a compact binary ($\text{binary}_{2-3}$). The second and third stages demonstrate the Roche-lobe filling of the triple system, and the triple common envelope, respectively.}
\end{figure}

A circumstellar TCE occurs when a more-compact binary (hereafter $\text{binary}_{2-3}$ as presented in Figure \ref{fig:systemscatch}  , or the
inner-binary, as it is termed in the context of hierarchical triples)
orbits an evolved star which fills its Roche-lobe (see middle panel of Figure \ref{fig:systemscatch}). Then, similarly
to the binary CE case, if the mass-transfer is unstable, a shared
envelope is formed. The evolution in this case will be somewhat different
due to the potential additional energy input from the $\text{binary}_{2-3}$
system and/or the more complex and potentially chaotic dynamics of
the (possibly unstable) triple. Moreover, the motion and evolution of $\text{binary}_{2-3}$ could be affected by the interaction with the
giant's gaseous envelope. As we discuss in the following, the outcome of such process can be a merger of the giant's core with one or more of the companions, a merger of $\text{binary}_{2-3}$'s components, or a
tidal disruption of the binary and the possible ejection of one of
the stellar components. The different evolution could also induce a different
structure of the remnant planetary nebulae around the resulting system
\citep{1992AJ....104.2151S}.

A TCE could also involve a different, circumbinary configuration.
In this case a binary CE is formed in a system with a distant third
companion, i.e. the evolved star is now part of an inner binary,
orbited by an outer third component. The expansion of the envelope
during the binary CE may then lead to engulfment of the third companion,
and its potential in-spiral onto the binary in a TCE. Here we focus
on the circumstellar case; the circumbinary case will be explored
elsewhere. 

Past hydrodynamical simulations of binary CEE encountered major difficulties;
they show that following the in-spiral phase most of the envelope
is not ejected, but only expands to larger size while
still remaining bound to the binary (e.g. \citealt{Ric+12,Iva+13,Iva+15,Kur+16,Ohl+16,Iac+17}). Consequently, other physical processes  have been suggested as possible causes for its ejection; these include jets launching inside the CE \citep{2019MNRAS.488.5615S, 2019MNRAS.490.4748S}, long period pulsations \citep{Cla+17} and dust driven winds \citep{DustDrivenWindsPaper}. Other suggested the importance of including the recombination energy (\citealt{Iva+15}), but this is doubtful as being the sole reason for the ejection- \citep{2018MNRAS.478.1818G} as well as for explaining formation of wide post-CE binary systems \citep{Reichardt2020}. In this paper we are not aiming to solve those problems, but only on the comparison between the outcomes of the CEE of binary and triple systems.

The paper is structured as follows. We first describe our simulation
methods in the following section. We then present our results (section
\ref{sec:Results}) and possible outcomes of circumstellar TCE, discuss
them in section \ref{sec:Discussion}, and summarize.

\section{Methods}
In our models we simulate the dynamical evolution of an evolved triple
system in which a binary system orbits a more massive red giant (which
we term the circumstellar case). The giant's envelope engulfs the
binary, leading to the binary's in-spiral into the giant envelope
and producing a triple common envelope (TCE) configuration. For this purpose,
we first modeled a red-giant star using the MESA 1D stellar evolution
code \citep{2011ApJS..192....3P} (version 2208), and then mapped its density profile
into a 3D model to be used in the hydrodynamical code GADGET2 \citep{2005MNRAS.364.1105S}.
After relaxing the star in the hydrodynamical code, we coupled the
star to an outer-orbiting binary ($\text{binary}_{2-3}$; modeled as two point mass stars,
unlike the fully hydro-modeled giant star, and coupled the hydro model
with a higher resolution dynamical code). 

We use the AMUSE - Astrophysical Multi-purpose Software Environment
\citep{2009NewA...14..369P} as a platform for coupling between several
external codes used for the physical processes. In particular, we used MESA
\citep{2011ApJS..192....3P}, the stellar evolution code, in order to produce
the initial profile of the RG star. The dynamics of the stars of $\text{binary}_{2-3}$ is modeled by the dynamical code HUAYNO \citep{2012NewA...17..711P}
which is a N-Body code, and the gravitational influence of $\text{binary}_{2-3}$ on the
envelope is modeled by MI6 \citep{10.1093/pasj/59.6.1095}. In the
following we provide a more detailed description of the different
parts of our model.

\subsection{Modeling the giant gaseous envelope and core}

\subsubsection{Stellar evolution}
\begin{figure*}
\includegraphics[width=\textwidth,clip]{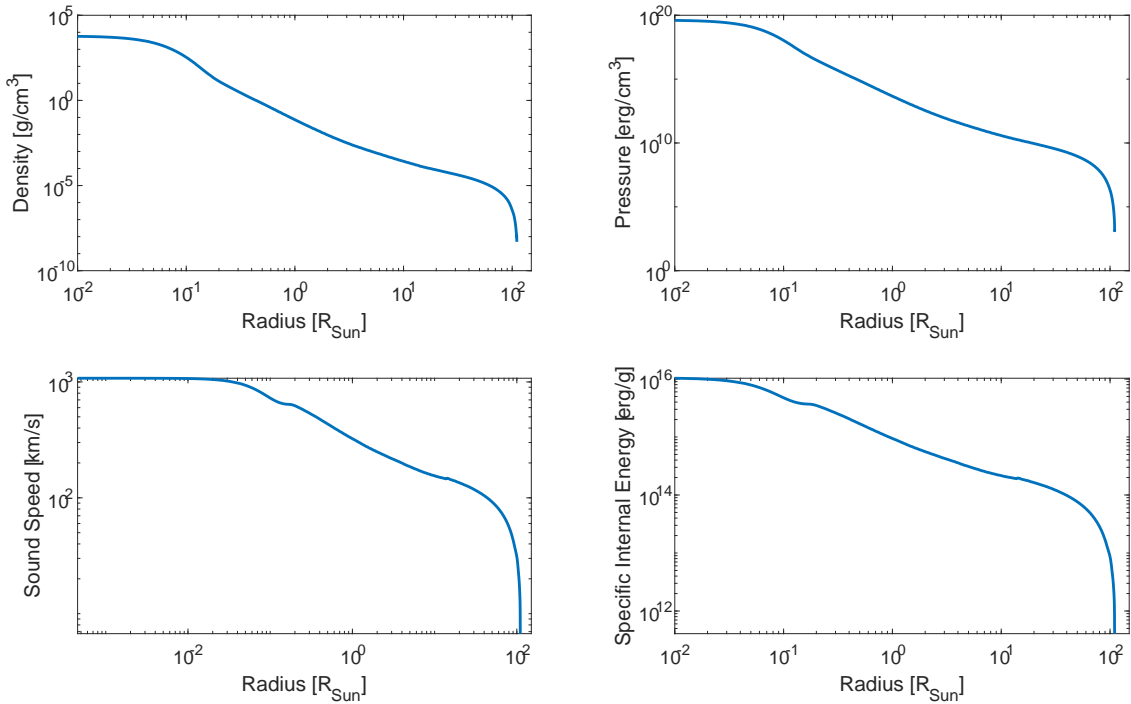}
\par
\caption{\label{fig:prerelaxprofiles}Radial profiles of the evolved  $8M_{\odot}$ stellar model}
\end{figure*}

The initial giant model is created with the stellar code MESA \citep{2011ApJS..192....3P}.
We create the initial model by simulating the stellar evolution of a
star from the zero-age main sequence stage up to the red-giant stage.
Here we explored cases where the giant reached its maximal radius
(if it were to evolve in isolation) when the TCE stage ensues. We
run MESA within the AMUSE framework. This allows
us to better control the stopping condition by running the evolution
step by step. Furthermore, the final model can then be simply coupled
with the other few-body modeling components discussed above. 
We considered two possible masses for the primary red-giant, an $8M_{\odot}$ case and a $2M_{\odot}$ case. In both cased we evolved the primary star up to the red giant (RG) phase. The giant with $8M_{\odot}$ was evolved until it reached a radius of  $R_{\star}\approx0.5AU\approx110R_{\odot}$. Its hydrogen- exhausted
core is approximately $M_{\text{core}}\approx1.03M_{\odot}$ with a
core radius of $R_{\text{core}}\approx0.09R_{\odot}$.  Its dynamical time scale is $\approx5.5$ days and the thermal time scale- $\approx5160$ years. The radial profiles of the evolved model are presented in Figure \ref{fig:prerelaxprofiles}.
The $2M_{\odot}$ giant was evolved until it reached a radius of $R_{\star}\approx50R_{\odot}$. Its 
core-size was found to be approximately $M_{\text{core}}\approx0.36M_{\odot}$ with a
core radius of $R_{\text{core}}\approx0.015R_{\odot}$ (similar to the model simulated by \citealt{Ohl+16}). Its dynamical time scale is $\approx3.3$ days and the thermal time scale- $\approx4500$ years.

\subsubsection{Smoothed particle hydrodynamics simulation}
We use the GADGET2 \citep{2005MNRAS.364.1105S} smoothed-particle
hydrodynamics (SPH) for the hydro-dynamical modeling. The SPH method
works by dividing the fluid into a set of discrete elements, referred
to as particles. These particles have a spatial distance, termed the
 \textit{"smoothing length"}, typically represented
in equations by $h$, over which their properties are "smoothed"
by a kernel function. This means that the physical quantity of any
particle can be obtained by summing the relevant properties of all
the particles which lie within the range of the kernel, and the contribution
of each particle is weighted according to its distance from the particle
of interest. 

In GADGET2, the kernel function is \citep{1992ARA&A..30..543M}:
\[
W\left(r,h\right)=\frac{8}{\pi h^{3}}\begin{cases}
1-6\left(\frac{r}{h}\right)^{2}+6\left(\frac{r}{h}\right)^{3} & 0\leq r\leq\frac{h}{2}\\
2\left(1-\frac{r}{h}\right)^{3} & \frac{h}{2}\leq r\leq h\\
0 & h<r
\end{cases},
\]
where r is the relative distance between the two particles and h is the smoothing length of the particle.  One should note that the smaller is the smoothing length, the
smaller is the number of neighbors that should be taken into account
for the calculation of each SPH particle. Another important parameter
is the softening length, which keeps the simulation from non-physical
behavior at very small separations between the point-mass particle and other particles, either point-mass or SPH particles. The gravitational
potential of such particles is then
\[
\phi=\frac{GM}{\left|\Delta r^{2}+\epsilon^{2}\right|^{\frac{1}{2}}},
\]
where $M$ is the mass of one point-mass particle, $\Delta r$ is the distance between the particle and the given location, and $\epsilon$ is the softening length.
In cases (like ours), where the simulation uses the softening
length for smoothing the kernel, $\epsilon$ can be thought as the radius
of the point-mass particle.

\subsubsection{Mapping the stellar model into the SPH model}
\paragraph{Mapping:}
The core of a giant star is much denser than its outer envelope,
and is not resolved in our simulations. Our focus here is not on modeling
physical mergers and interactions with the small and unresolved dense
core, but rather exploring the interactions with the stellar envelope.
We therefore represent the core in our simulations as a point mass particle(called "dark matter particle" in Gadget2, because of their application to cosmological simulations) without considering changes in its internal structure. We
chose the softening length such that the potential of the core is approximately its analytical form at distances larger than $2.8\epsilon$, $\epsilon\approx10r_{c}$, where $r_{c}$ is the radius
of the hydrogen-exhausted core, inducing the potential of a Plummer sphere of size $\epsilon$.

To convert the stellar model created by MESA into a 3D SPH model,
we used AMUSE's function Star\_to\_sph \citep{AMUSE-Book}. This function
converts the core region into a point mass particle with an effective radius as described above, and corrects the density and internal energy profiles to maintain pressure equilibrium and conserve the original entropy profile (as explained in \citealp{deV+14}). It then divides this external region to our desired number of particles, with equal masses
and different smoothing lengths. Each gas particle has its own gravitational potential
and can interact with its surrounding, i.e. we use the Lagrangian
form of the fluid equations of motion. 

\paragraph{Relaxation:}
Following the mapping of the 1D model from MESA to 3D, and the use
of somewhat different equation of state between the codes (the 1D radial OPAL EOS in MESA, which is explained in \citealt{2011ApJS..192....3P} and 3D hydrodynamical equations in Lagrangian form in Gadget2, as described in \citealp{2005MNRAS.364.1105S}), a relaxation
stage is required as to obtain a stable stellar configuration for
the SPH model. During this stage, we keep the center of mass (COM) position and COM
velocity constant; we adiabatically adjust the position and velocity
of each SPH particle for $130\text{ days} \approx24$ dynamical times (for our more massive giant). The particle positions and velocities are adjusted after each step in the following way
\[
r_{i,j}=\left(r_{i,j}-r_{COM,i}\right)+r_{COM,0}
\]
\[
v_{i,j}=\left(v_{i,j}-v_{COM,i}\right)\cdot\left(i/nsteps\right)+v_{COM,0},
\]
Where $r_{i,j}$ and $v_{i,j}$ are the j'th gas particle position and velocity at step $i$, $r_{COM,i}$ and $v_{COM,i}$ are the giant's center of mass position and velocity at step i, $nsteps$ is the total number of the damping steps. After each step the internal velocities are damped by a factor which decreases from 1 in the first step to 0 in the last one.

We first test the stability of our massive giant star model in isolation. For this purpose, we used a similar method as described in \citet{Iac+17}. Following a damping phase as described above, we continued the evolution of the giant for an additional $130$ days (same amount of dynamical times) with no damping. At each time step we compared the gas particle velocities with some typical velocity scales: the local sound speed, $c_s$, the initial velocities of $\text{binary}_{2-3}$'s components, relative to the center of mass of the giant, and the escape velocity, $v_{esc}=R_{\star}/t_{dyn}$, where $R_\star$ and $t_{dyn}$ are the initial star radius and dynamical time, respectively. During those $\sim24$ dynamical time scales, at most $0.05$ percent of the particle exceeded the lowest of those velocity limits. 

Since the star is stabilized in isolation, we could add the gravitational field of $\text{binary}_{2-3}$ during the damping phase and form a binary (trinary) system which is less affected by the sudden change in the potential, and by the change of the simulation code. More precisely, We follow the same method as mentioned above, but we now include the gravitational potential of $\text{binary}_{2-3}$. We do
not allow the binary to evolve during that time(i.e, it is considered as a constant
potential at this stage), where we follow the same approach as \citet{deV+14}.

While taking the companion into account during the relaxation
stage, we place it far enough from the giant such that it won't
considerably affect the giant's shape (nor initiate a RLO), but sufficiently near such that
the system will eventually enter the CE stage during the simulation time. The result of such configuration is a "gradual" addition of the gravitational potential of the companion, which is more natural than its sudden inclusion where it has a greater effect.

\begin{figure}
\includegraphics[width=\linewidth]{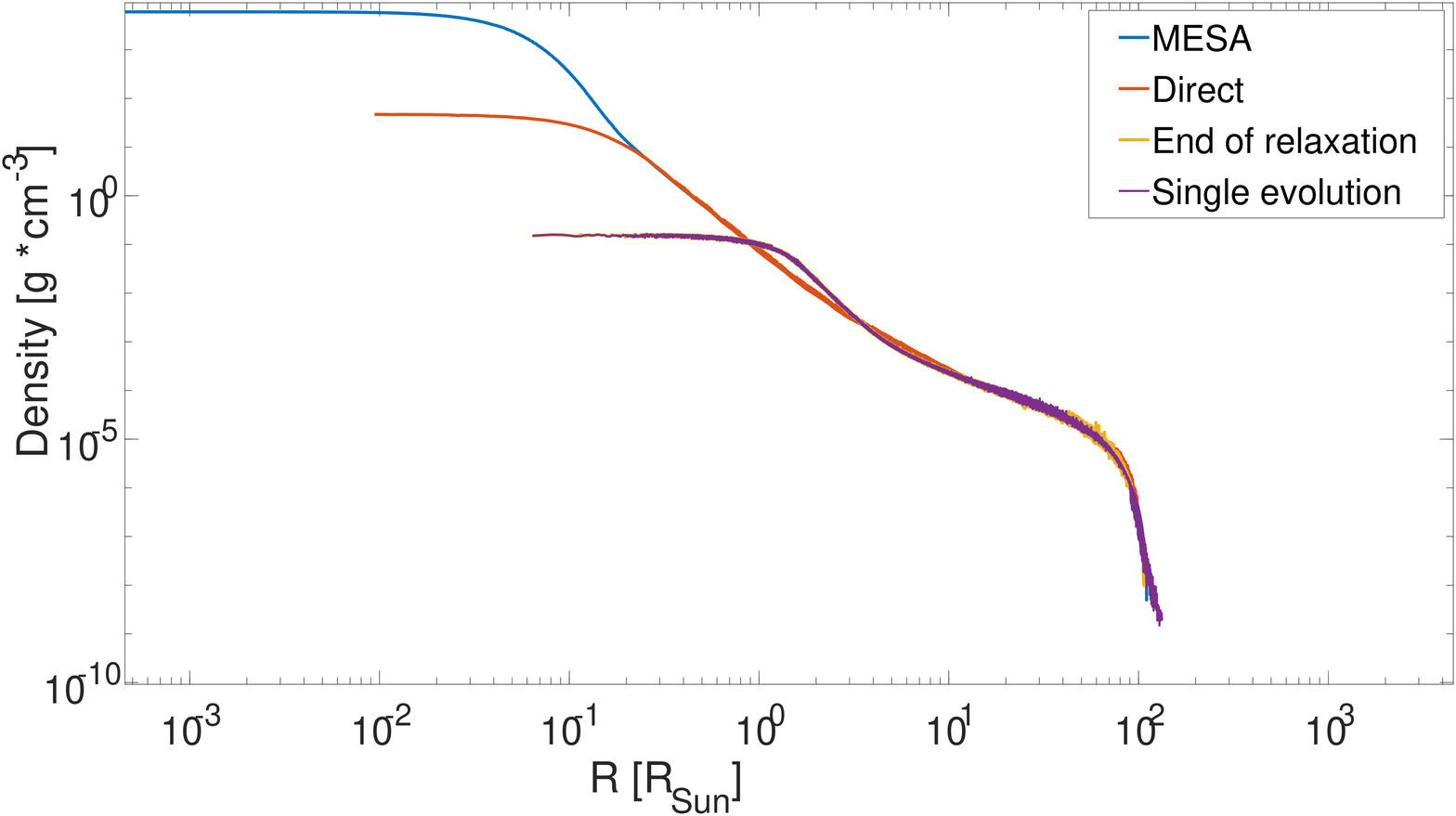}
\caption{\label{fig:relaxedprofiles} Radial density profiles  of the evolved  $8M_{\odot}$ stellar model, before mapping to a 3-D model (blue); immediately following the mapping (orange); after the damping phase (yellow); and following the later evolution in isolation (purple). }
\end{figure}

\subsection{Coupling of the red-giant with the inner (point masses) \texorpdfstring{$\text{binary}_{2-3}$}{}}
The motion of the system of the $\text{binary}_{2-3}$ is a classical 2-body
problem, which is perturbed by the gravitational potential of the
core and the envelope of the giant star. In order to improve the model
accuracy of the binary orbital motion and the momentum conservation,
we make use of the HUAYNO code \citep{2012NewA...17..711P}, which
is a special class of Individual Time-step Scheme, semi-symplectic
direct N-body integrator. The giant star gaseous envelope is modeled
using the SPH code Gadget2 as discussed earlier. The N-body code is
coupled with the SPH code using the AMUSE environment. We effectively
split the Hamiltonian into the different parts, which are modeled using
the different codes. This combination can give us better approximation
for each of the sub-systems. The two sub-systems are then linked through
an AMUSE ``bridge'' between $\text{binary}_{2-3}$ and the gas using the
MI6 code \citep{10.1093/pasj/59.6.1095}; a 6th order N-Body integrator
with mixed 4th and 6th order Hermite integration scheme, originally
developed for simulating the galactic center. Like in the galactic
center, we have a multiple system ($\text{binary}_{2-3}$) affected by the
gravitational potential of the third star core and its gaseous envelope (calculated with Gadget2),
and affecting the envelope evolution. Overall we follow a similar
approach as used by \citet{deV+14} to study mass-transfer dynamics in
triple systems (that are not evolving through TCE phase). 

\subsection{Code bench-marking and resolution}
\begin{figure}
\includegraphics[width=\linewidth,clip]{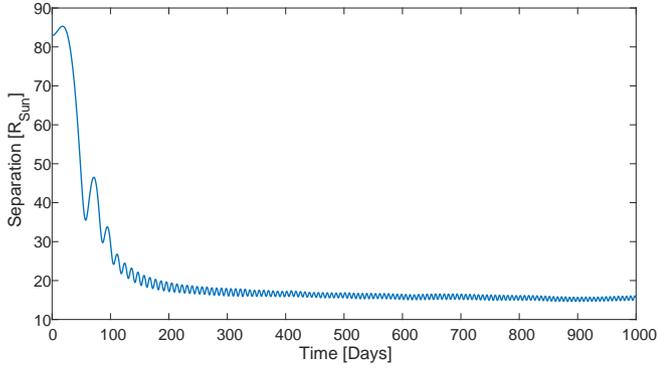}
\par
\caption{\label{fig:passy}  Simulation 16 (see \ref{tab:system-configurations}, with similar configuration to \citealt{Pass+12}, of $1 M_\odot$ giant with $0.6 M_\odot$ companion. This simulation was done in order to benchmark and test the use of Gadget2 \citep{2005MNRAS.364.1105S} for fully simulating common envelope evolution, not done previously, to the best of our knowledge. We find our simulation well reproduces previous simulations done with other codes (compare with \citealt{Pass+12}), showing similar evolution timescales and kinematics, both in the rapid inspiral stage and later during the slower evolution, and giving rise to similar final separations.}
\end{figure}

\begin{figure}
\includegraphics[width=\linewidth,clip]{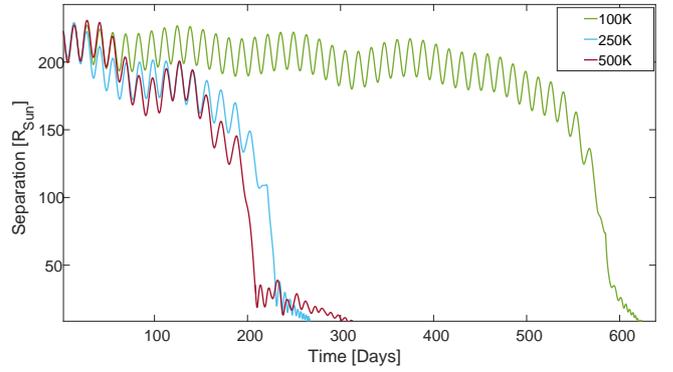}
\par
\caption{\label{fig:Resolution}  Separation between the giant's core and the companion which will finally merge in simulation 1 (see Table \ref{tab:system-configurations}) that was ran with 3 different resolutions, i.e- different number of SPH particles. We can see that the runs with 250K and 500K particles have relatively similar initiation time of the fast plunge-in and in the overall duration until merging with the core. The plunge-in phase began much later in the run consisted of 100K particles, but then had a faster duration. Overall, the outcomes where the same in all of those simulations- one of the companions merged with the core, while the other one remained bound at a closer distance than prior to the CE, but did not merge. }
\end{figure}

In order to test our code, we first successfully reproduced the simulated
evolution of previously studied binary common envelope and triple
mass-transfer systems (\citealt{Pass+12}-  see Figure  \ref{fig:passy} , and \citealt{deV+14}, respectively).
We carry out resolution tests in the limits of long computational times, with several TCE models, at progressively higher resolution, up to 500K.
We found that the use of 250K SPH particles produced similar results
compared with higher resolution simulations (500K) and we use this
resolution throughout the models discussed below (See Figure \ref{fig:Resolution} ). 
Our models with 100K, 250K and 500K show consistent results. Nevertheless,  higher resolution simulations are desired in order to confirm the convergence. Given the high computational cost needed for higher resolution simulations, we resort to the current resolution allowing us to model a reasonable phase space, of models with our available computational power. These resolutions compare well with most currently-run cutting-edge simulations of CEE.

All models were running on the Astric computer cluster of the Israeli I-CORE center, with up to 32 cores used in each run. Even with these computational resources, each of the TCE simulations with more than 250K particles required over than a month to run.

\begin{figure}
\includegraphics[width=\linewidth,clip]{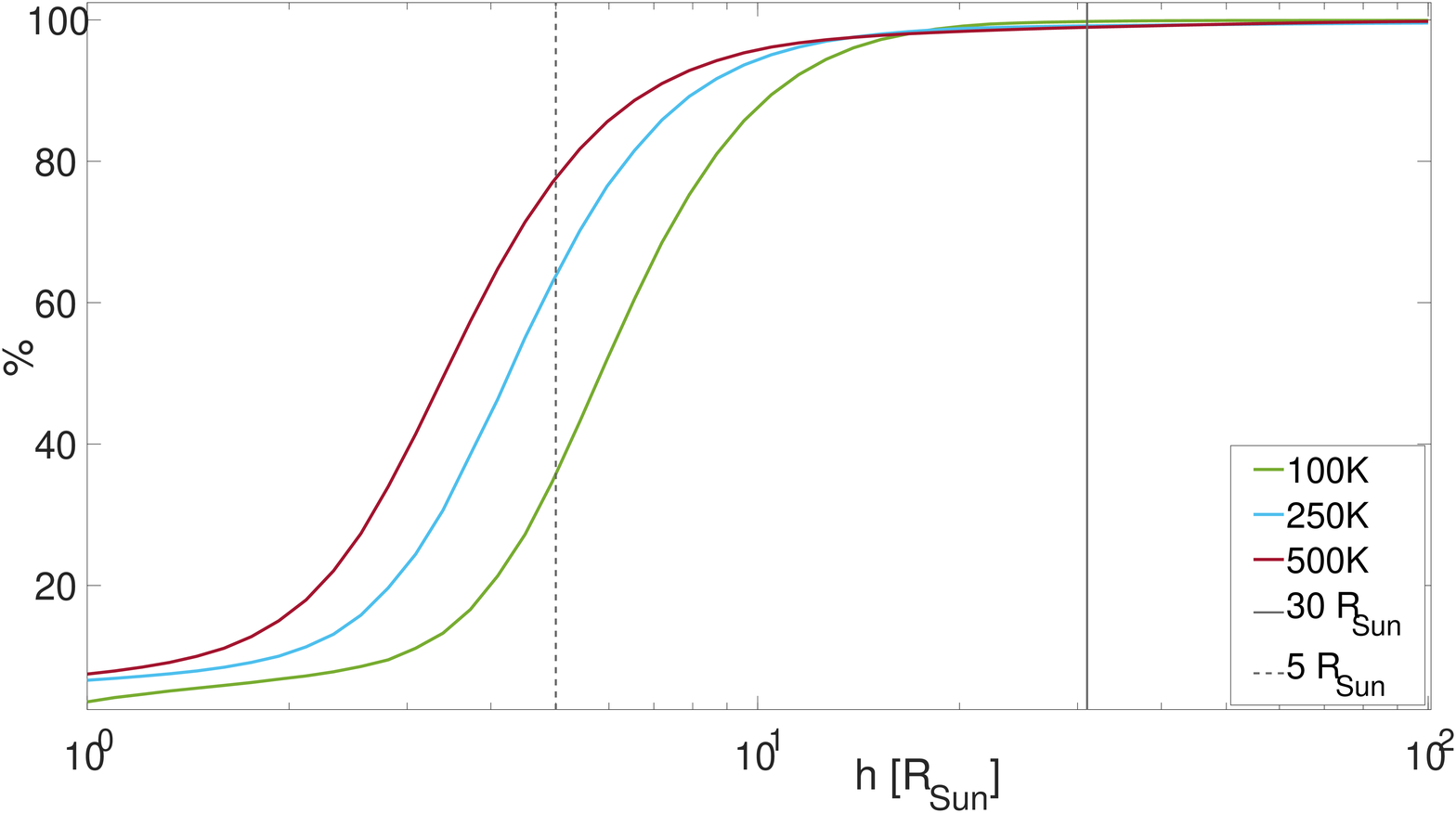}
\par
\includegraphics[width=\linewidth,clip]{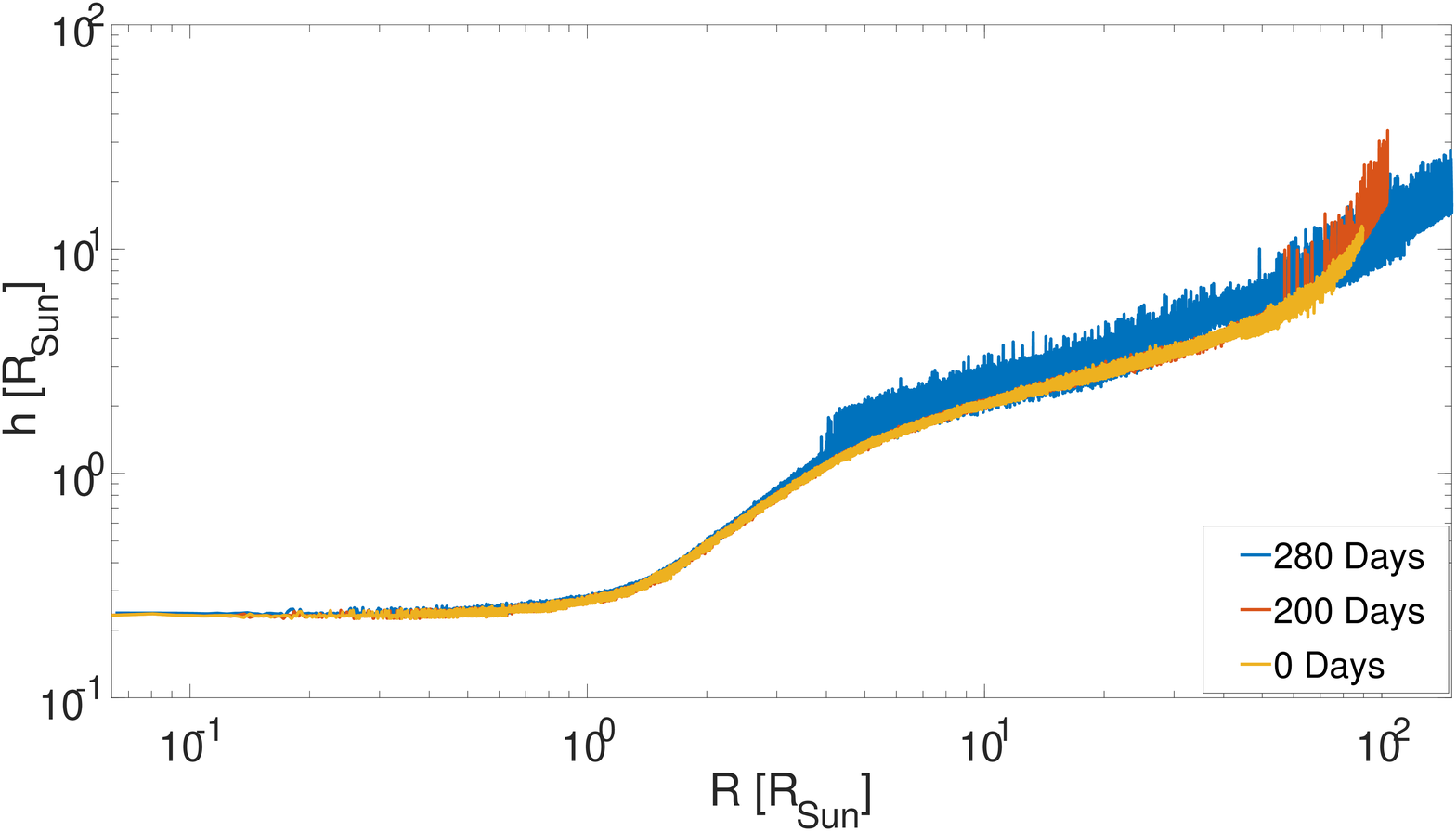}
\caption{\label{fig:hcompare}  Upper Panel: Comparison between the different smoothing length distributions at  different resolutions. The profiles were taken approximately at the beginning of the fast plunge-in. The vertical lines represent the approximate size of $\text{binary}_{2-3}$, for a comparison with the gas smoothing lengths.
Lower Panel: The resolution during different times of the simulation. Most of the changes in the smoothing lengths occur only at the outer parts of the expanding envelope.}
\end{figure}

We examined the resolution of our models by comparing the histogram of the different smoothing lengths throughout the envelope in our model (simulation 1 in Table \ref{tab:system-configurations}), and its distribution at different simulation times (see Figure \ref{fig:hcompare}). Since most of the SPH particles in this resolution had much smaller smoothing length than the orbit of the $\text{binary}_{2-3}$, the gas interaction can affect $\text{binary}_{2-3}$ even in the outer region, where the resolution is the lowest (see bottom of same Figure \ref{fig:hcompare}). In addition, the distribution of the smoothing lengths in the dense region, close to the core, varies only slightly during the simulation. Since our current focus is not the mass ejection from the outer layer, we can neglect those resolution variations.
The components of $\text{binary}_{2-3}$ are modeled as point masses, and have no softening length affecting their mutual interaction (modeled with HUAYNO code), but their interaction with the gas and the stellar core is smoothed with $\epsilon\approx 10 r_c$, the same value as that of the core particle (which has a similar mass). We should note, however, that due to the relatively close values of the compact binary separation and the average local smoothing length of the gas, this resolution is not sufficient for an explicit determination of the boundary between initial separations of $\text{binary}_{2-3}$ that lead to their mutual merger, and those that end with their disruption.

\begin{figure}
\includegraphics[width=\linewidth,clip]{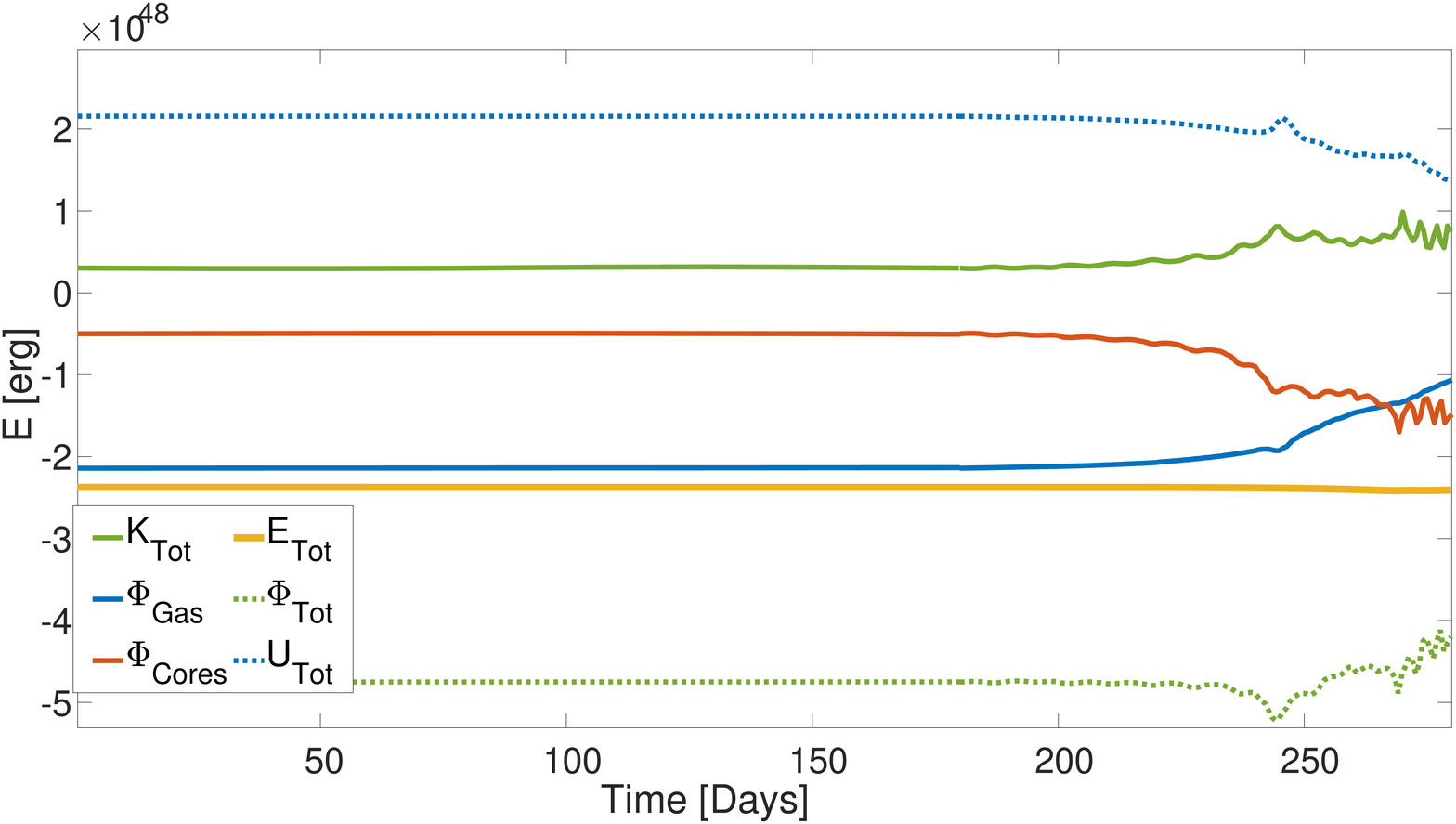}
\par
\includegraphics[width=\linewidth]{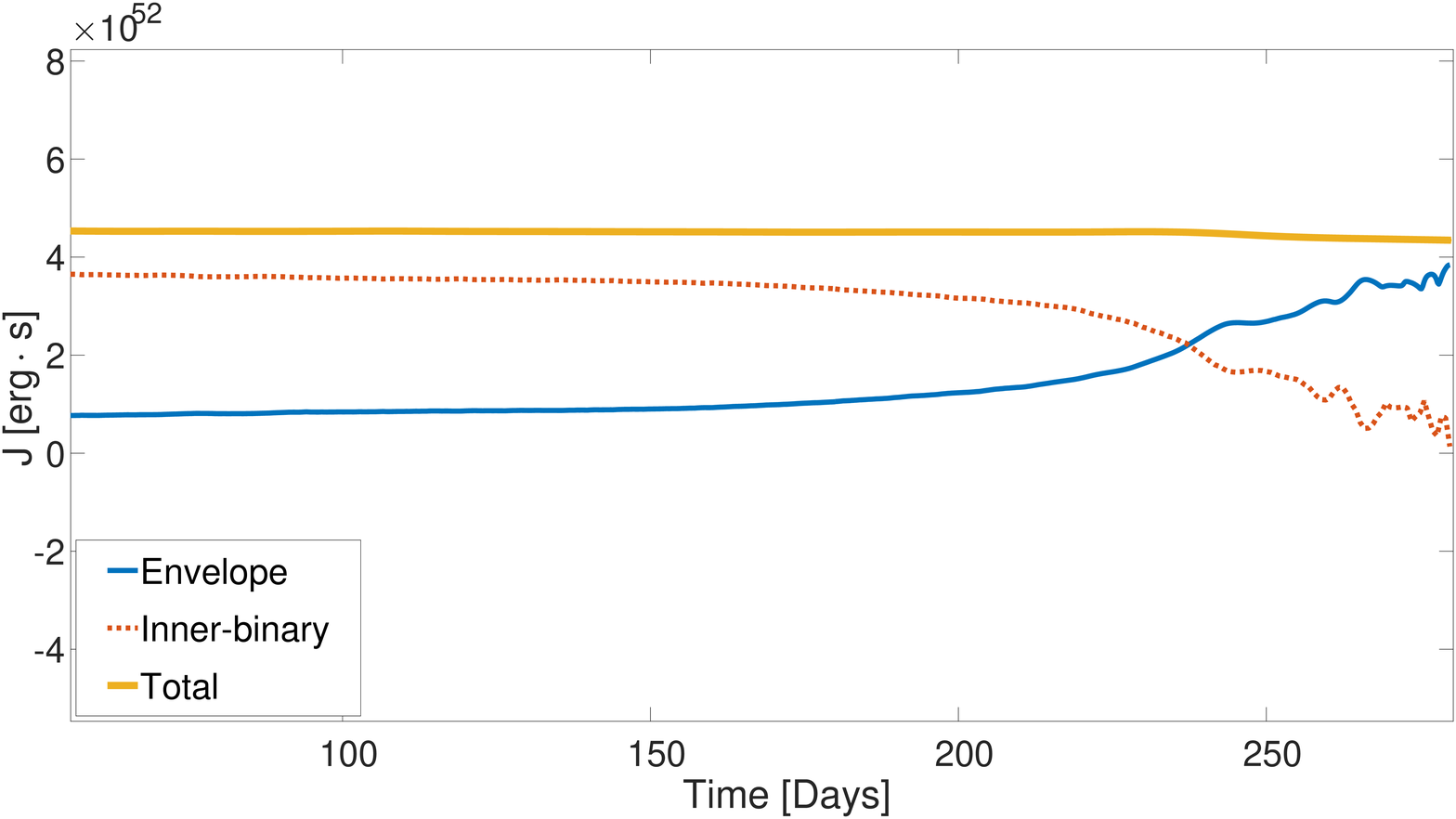}
\caption{\label{fig:conservation} Energy and angular momentum conservation. 
\newline
Upper Panel: Energy conservation as a function of time (for model 1 in Table \ref{tab:system-configurations}). The potential energy of the outer binary ($\text{binary}_{1-23}$ in Figure \ref{fig:systemscatch}) increases while the gas potential energy of the expanding envelope decreases, as well as its internal energy. The gas internal energy is the total internal energy of the system, since the components of $\text{binary}_{2-3}$ and the giant core are modeled as "point-mass" (gravitational only) particles. The self potential energy of $\text{binary}_{2-3}$ is not presented here due to its relatively small value (and thus change) during the run, in comparison with the red-giant potential. The kinetic energy at the beginning of the simulation is dominated by the energy of $\text{binary}_{2-3}$, but later on the gaseous envelope gains some velocities in the outwards direction that leads to its partial unbinding.\newline
Lower Panel: Angular momentum as a function of time. The dotted red line shows the total angular momentum of $\text{binary}_{2-3}$ (sum of the angular momenta of both components), the solid blue line presents the envelope's angular momentum, and the total angular momentum of the system is shown by the thick yellow line.}
\end{figure}

Figure \ref{fig:conservation} shows that our simulation conserved energy and angular momentum to less than 1 percent.
The bottom panel shows the angular momenta components relative to the center of mass of the entire system. As expected from the onset of the CE, the angular momentum of the outer binary ($\text{binary}_{1-23}$ , i,e - $\text{binary}_{2-3}$ components' center of mass with the giant) decreases and the envelope's angular momentum increases as a consequence.

The duration of the relaxation stage and the CE stage were chosen
 to be 24 times larger than a dynamical time of 5.5 days. After a few simulations using
sink particles (to measure the accreted mass on a particle), accretion on the companion and core was found to be
negligible compare to their mass, in agreement with \citet{Pass+12}.
For that reason, we neglect any accretion effects. Some studies suggest
feedback from accretion outflows/jet may affect the evolution \citep[e.g][and references therein]{2017MNRAS.472.4361S,2019MNRAS.490.4748S,2020arXiv200204229S};
the potential importance of such effects, is still debated
and are beyond the scope of the current study.

Finally, we also compared the results of a TCE with a binary common
envelope evolution where $\text{binary}_{2-3}$ component in the TCE was
replaced by a single star of the same total mass, as to compare, at some level,
triple and binary common envelope evolution effects on systems of similar
masses and outer binary separations. 

 The different mass ratios between the evolved star and the companion(s) were chosen such that the corresponding binary CEs (simulations 14 and 15 in Table \ref{tab:system-configurations}) would allow us to explore two qualitatively different outcomes; one is a merger outcome (simulation 14), whereas the other (15) results in the formation of short period binary on a stable orbit. As we show below,  these final outcomes, either a core merger with one of the components or a stable orbit- do not change when replacing the companion with a binary as to form a TCE.

All simulations ran until a "merger" happened or the evolution reached $1400\,\text{days}$.  The condition for such a merger of two point mass particles is when their relative distance is shorter than the sum of their smoothing lengths \ radii, after this point our simulations are no longer applicable.

\section{Results}
\label{sec:Results}
\subsection{Triple common envelope outcomes}
We studied a limited grid (given the computational cost) of possible
configurations, and considered two different mass ratios. The first grid is for
a $8M_{\odot}$ RG with two $1M_{\odot}$ companions, where we varied
the following parameters with two possible values for each parameter,
and considered all the possible combinations for a total of 8 modeled
configurations. These parameters include (1) the initial separations;
(2) the relative inclination; and (3) the orbital phases (combined
with an additional $90^{\circ}$ orbital inclination, such that $\text{binary}_{2-3}$ orbits the giant on a plane perpendicular to the orbital plane of the giant with the compact binary, and both companions are initially located at the same distance from the giant's core); the model parameters
are listed in Table \ref{tab:system-configurations}. In addition,
we modelled three different inner separations of $\text{binary}_{2-3}$ for the second mass ratio
of $2M_{\odot}$(RG) and $0.6M_{\odot}$+ $0.4M_{\odot}$ companions.
Overall, we simulated 11 configurations for triple systems. As mentioned
above we consider only circumstellar configurations where a more compact,
point-mass binary orbits an evolved star and in-spirals in its envelope.
The summarized results of the evolutionary outcomes can be found in Tables \ref{tab:Results-summary-of820},\ref{tab:Results-summary-of8290},
and \ref{tab:Results-summary-of210}.

\begin{table}
\begin{tabular}{|c|c|c|c|c|c|}
\hline 
 No & Masses   &$a_{\text{out}}$& Inclination & $a_{\text{in}}$ & Orbital
\\ &(${\rm M_{\odot}}$)&(${\rm R_{\odot})}$ &&(${\rm R_{\odot})}$& phase\tabularnewline
\hline 
\hline 
1&8 + 1 + 1 &$216$& $5^{\circ}$ & $26$ & $0^{\circ}$\tabularnewline
2&8 + 1 + 1 &$216$& $5^{\circ}$ & $3$& $0^{\circ}$\tabularnewline
3&8 + 1 + 1 &$216$& $45^{\circ}$ & $26$ & $0^{\circ}$\tabularnewline
4&8 + 1 + 1 &$216$& $45^{\circ}$ & $3$ & $0^{\circ}$\tabularnewline
5&8 + 1 + 1 &$216$& $5^{\circ} + 90^{\circ}$ & $26$ & $90^{\circ}$\tabularnewline
6&8 + 1 + 1 &$216$& $45^{\circ} + 90^{\circ}$ & $26$& $90^{\circ}$\tabularnewline
7&8 + 1 + 1 &$130$& $5^{\circ} + 90^{\circ}$ & $26$ & $90^{\circ}$\tabularnewline
8&8 + 1 + 1 &$130$& $45^{\circ} + 90^{\circ}$ & $26$& $90^{\circ}$\tabularnewline
9&8 + 1 + 1 &$130$& $5^{\circ} + 90^{\circ}$ & $3$& $90^{\circ}$\tabularnewline
10&8 + 1 + 1 &$130$& $45^{\circ} + 90^{\circ}$ & $3$& $90^{\circ}$\tabularnewline

11&2 + 0.6 + 0.4 & $60$ & $0^{\circ}$ & $3$ &$0^{\circ}$\tabularnewline
12&2 + 0.6 + 0.4 & $60$ & $0^{\circ}$ &  $13$ &$0^{\circ}$\tabularnewline
13&2 + 0.6 + 0.4 & $60$ & $0^{\circ}$ & $26$ &$0^{\circ}$\tabularnewline

14&8 + 2 & $216$ &  \multicolumn{3}{|c|}{binary common envelope}  \tabularnewline
15&2 + 1 & $60$ & \multicolumn{3}{|c|}{binary common envelope}  \tabularnewline
16&1 + 0.6 & $83$ & \multicolumn{3}{|c|}{reproduced- \citealp{Pass+12}}  \tabularnewline
\hline 
\end{tabular}\caption{\label{tab:system-configurations}Initial values for the modeled systems,
where $a_{\text{in}}$ is the inner separation of $\text{binary}_{2-3}$. We varied the inner
binary separations for both mass models, and considered different
inclinations (between the orbital plain of $\text{binary}_{1-23}$ and $\text{binary}_{2-3}$), inner-binary ($\text{binary}_{2-3}$) orbital phase and outer-binary ($\text{binary}_{1-23}$) separation for the high mass TCE cases. We explored a total of 10 configurations
for the high mass-ratio case, and additional 3 cases for the low-mass case. In all models the evolved star is a Red Giant. The softening length of the core and  $\text{binary}_{2-3}$ components (with respect to the giant) are $\approx 0.9R_{\odot}$ for the high-mass cases and $\approx 0.15R_{\odot}$ for the low-mass cases.}
\end{table}

\begin{figure}
	
	\begin{tabular}{|c|c|c|}
	\hline
	 &face on  &edge on \tabularnewline
	\hline 
	\multicolumn{3}{|c|}{\includegraphics[width=0.415\columnwidth]{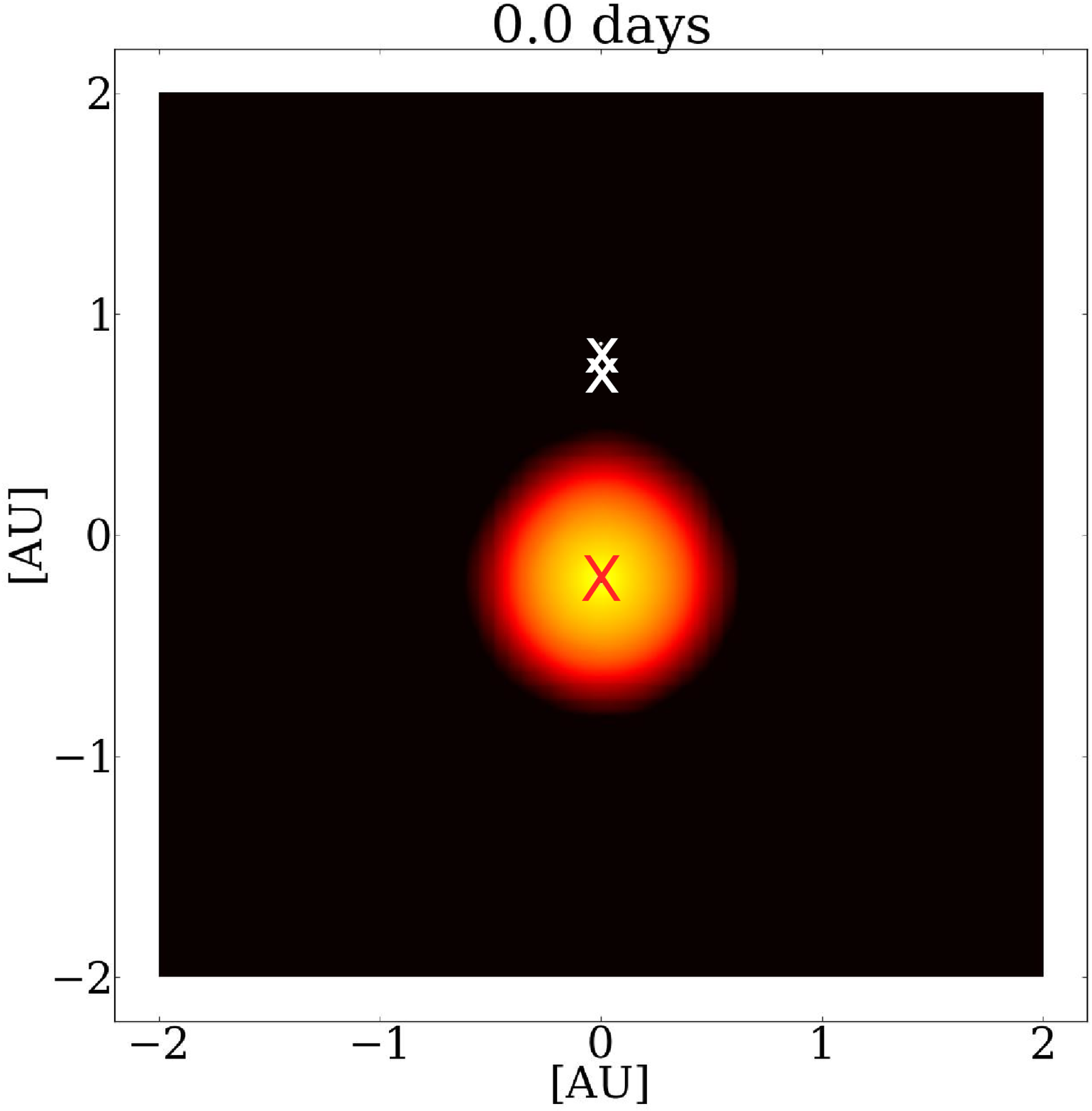}\includegraphics[width=0.4\columnwidth]{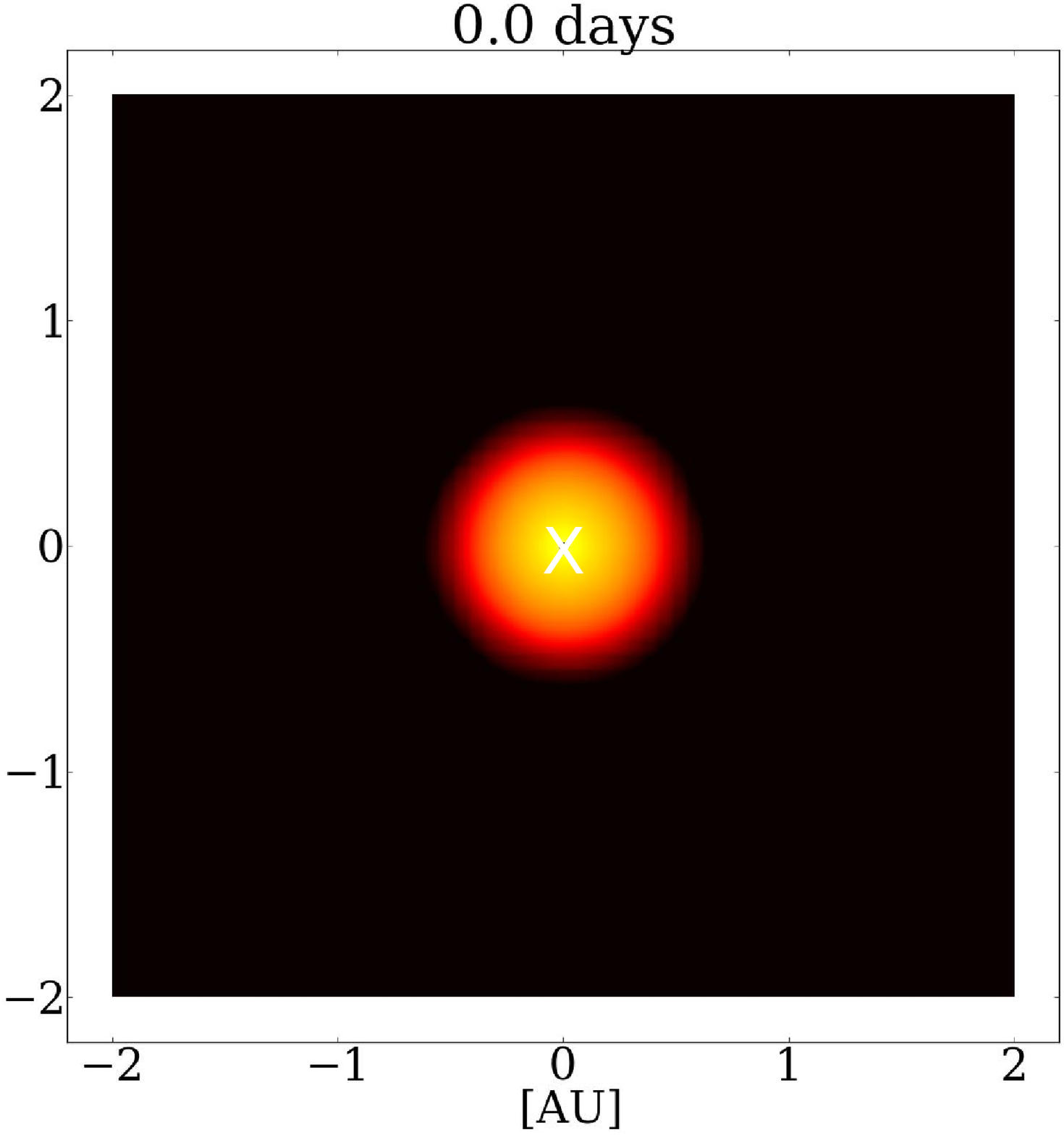} \includegraphics[width=0.09\columnwidth]{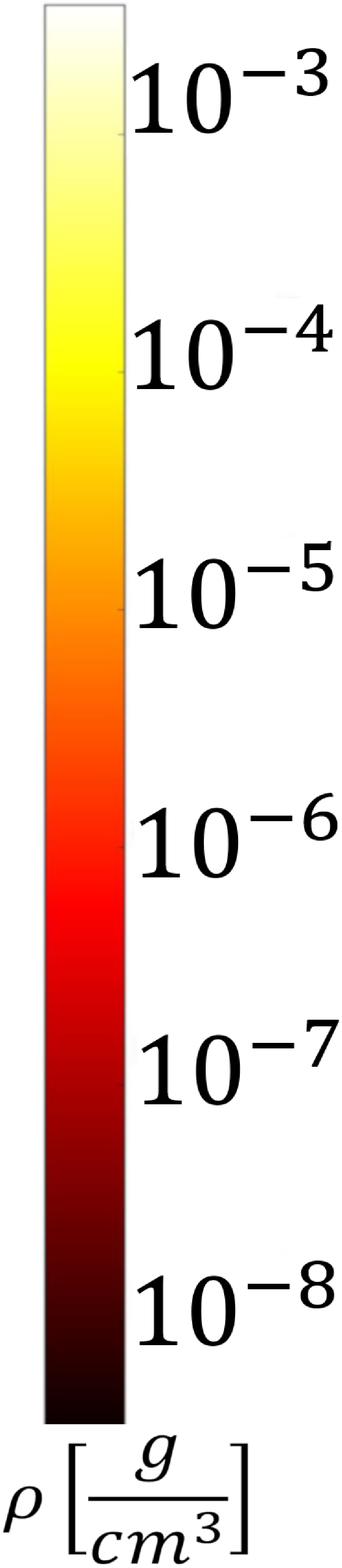}}\tabularnewline
	\multicolumn{3}{|c|}{\includegraphics[width=0.415\columnwidth]{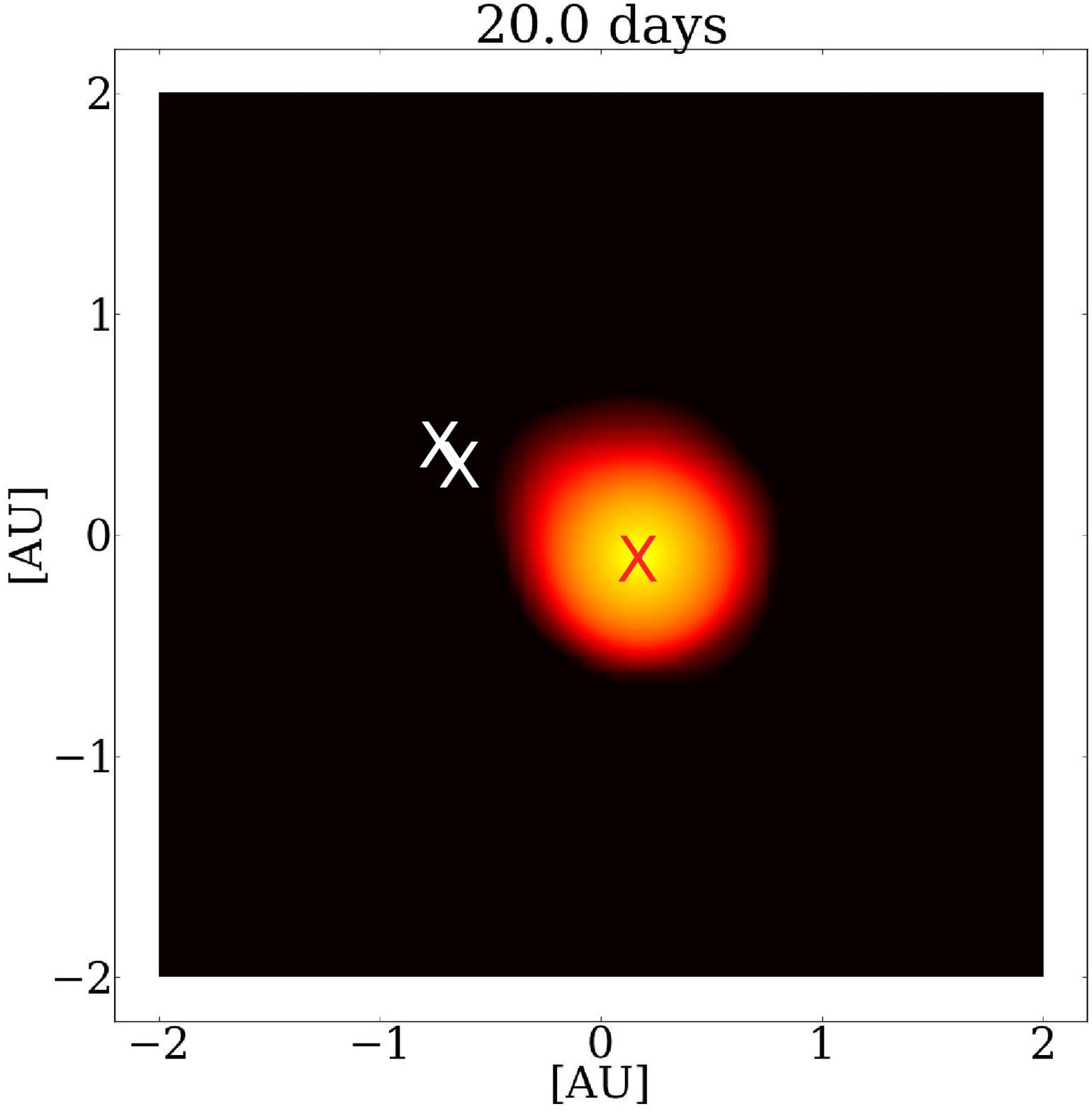}\includegraphics[width=0.4\columnwidth]{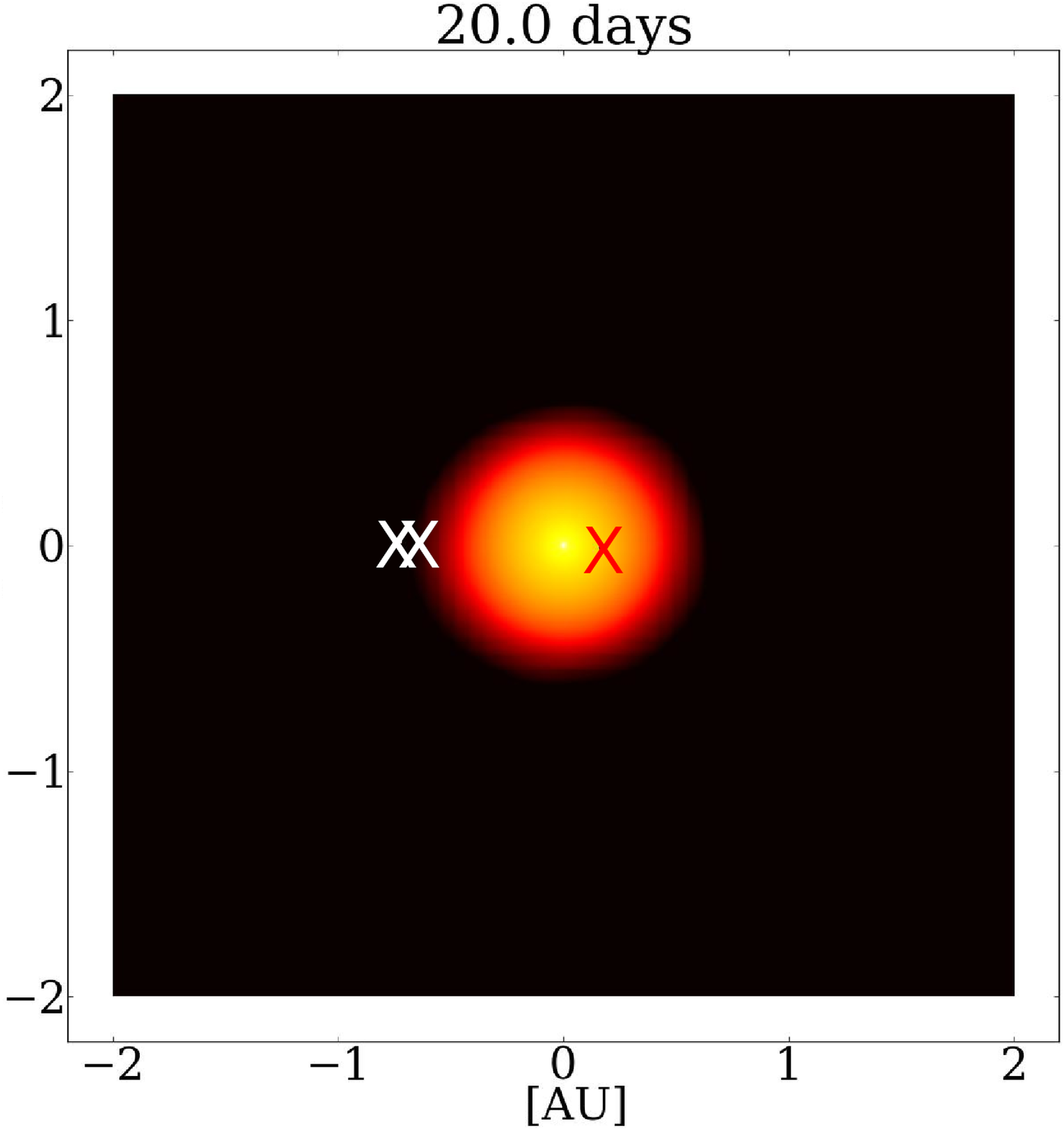}\includegraphics[width=0.09\columnwidth]{80265/colorbar}}\tabularnewline
	\multicolumn{3}{|c|}{\includegraphics[width=0.415\columnwidth]{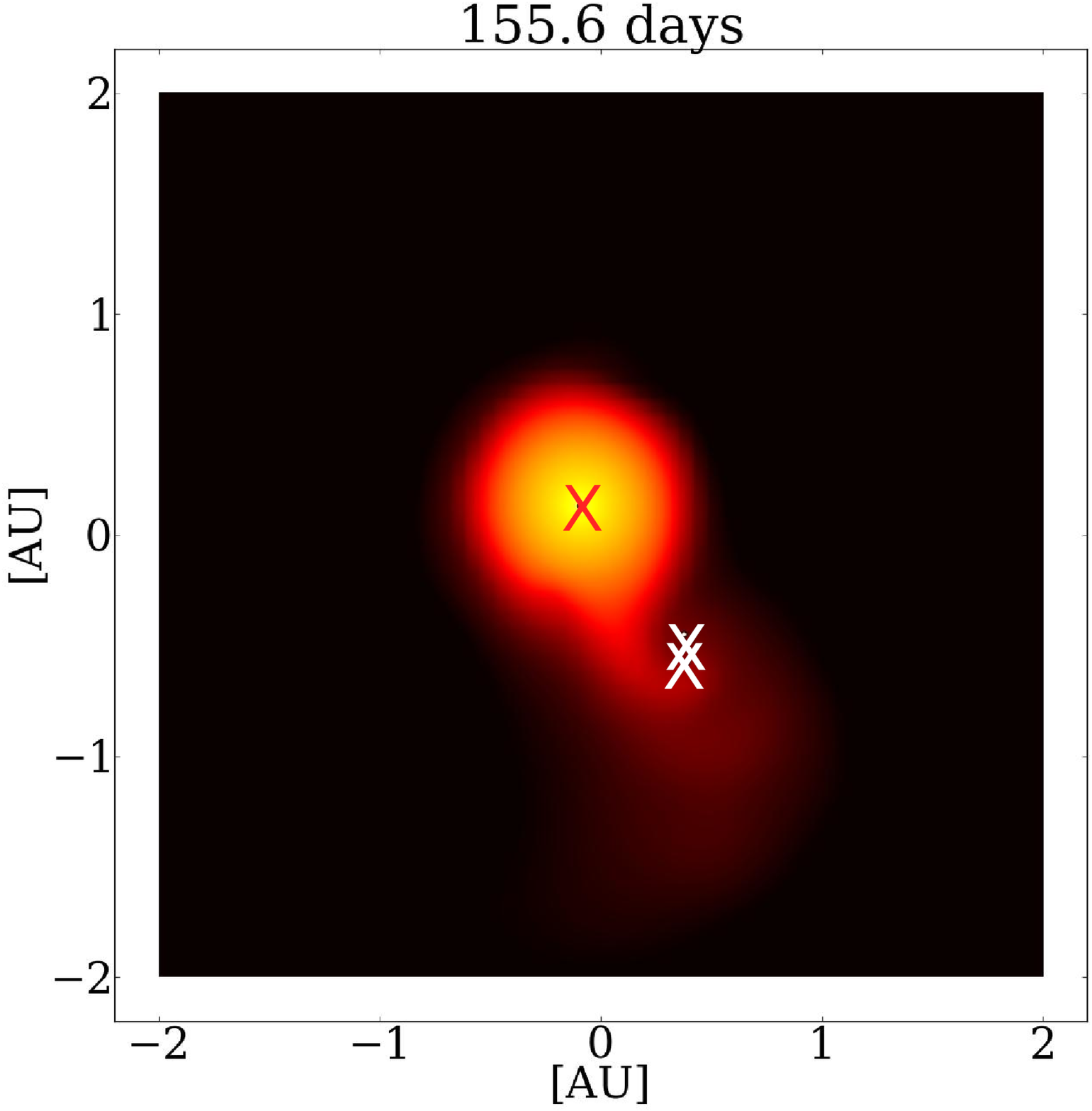}\includegraphics[width=0.4\columnwidth]{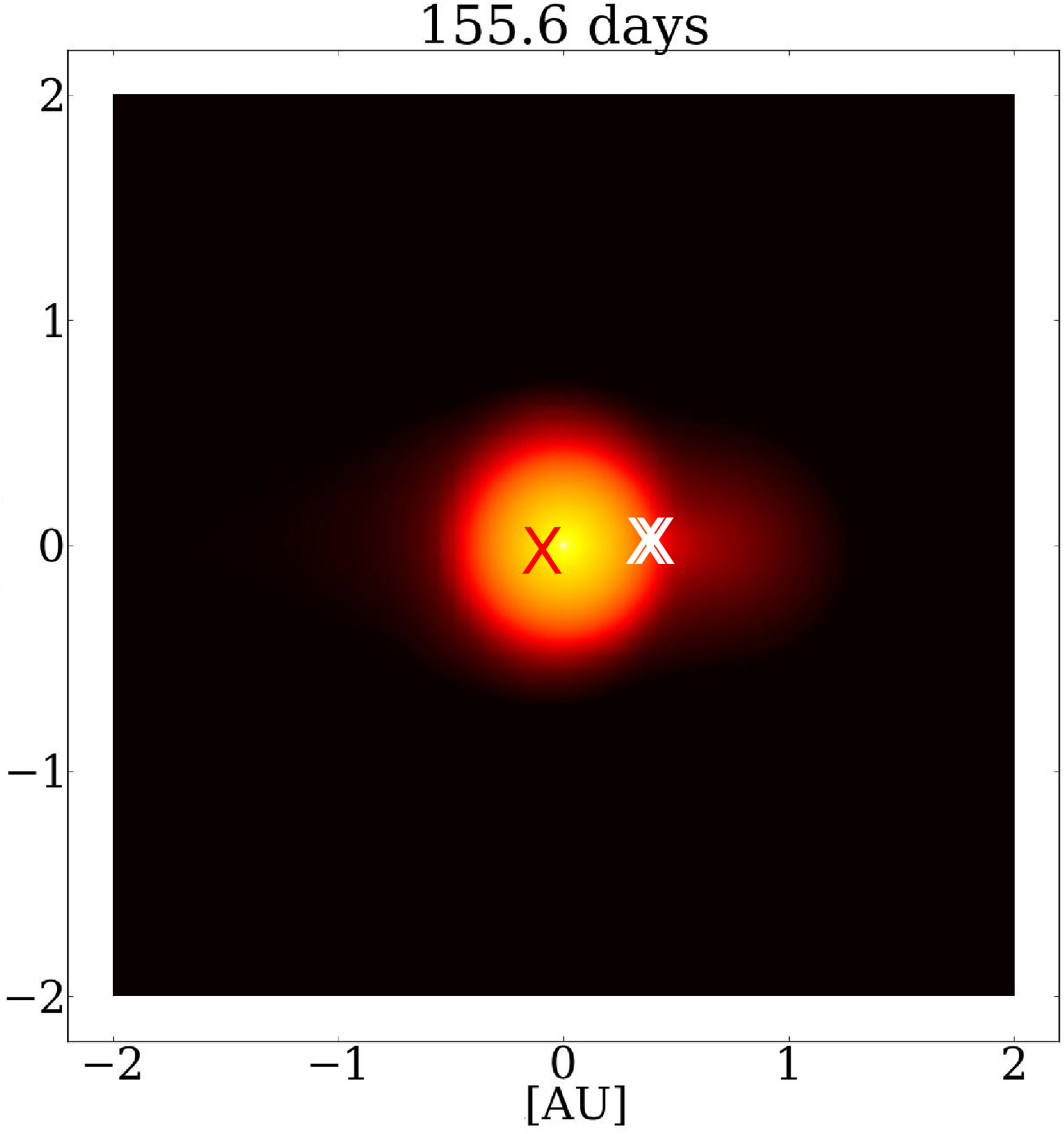}\includegraphics[width=0.09\columnwidth]{80265/colorbar}}\tabularnewline
	\multicolumn{3}{|c|}{\includegraphics[width=0.415\columnwidth]{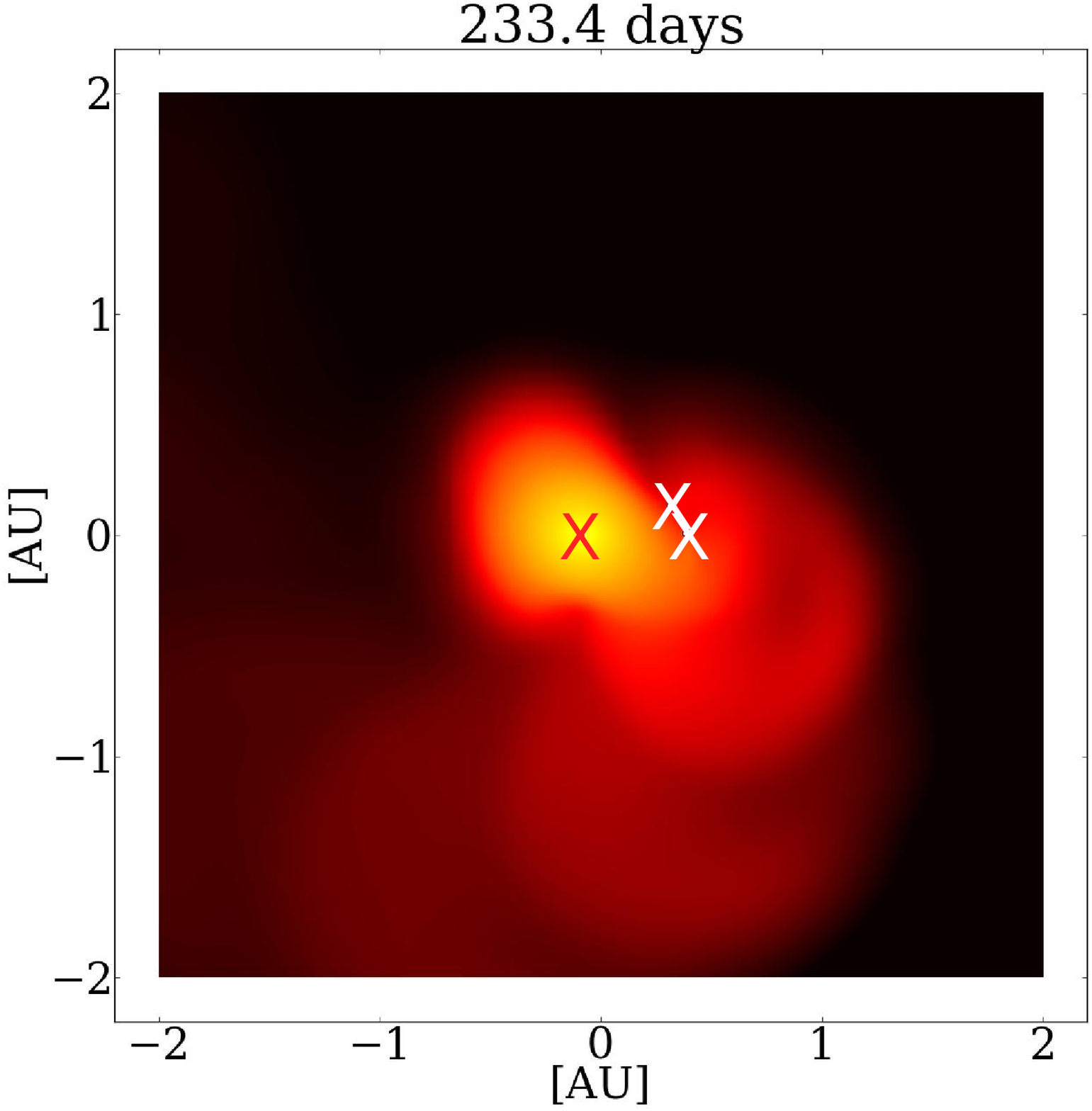}\includegraphics[width=0.4\columnwidth]{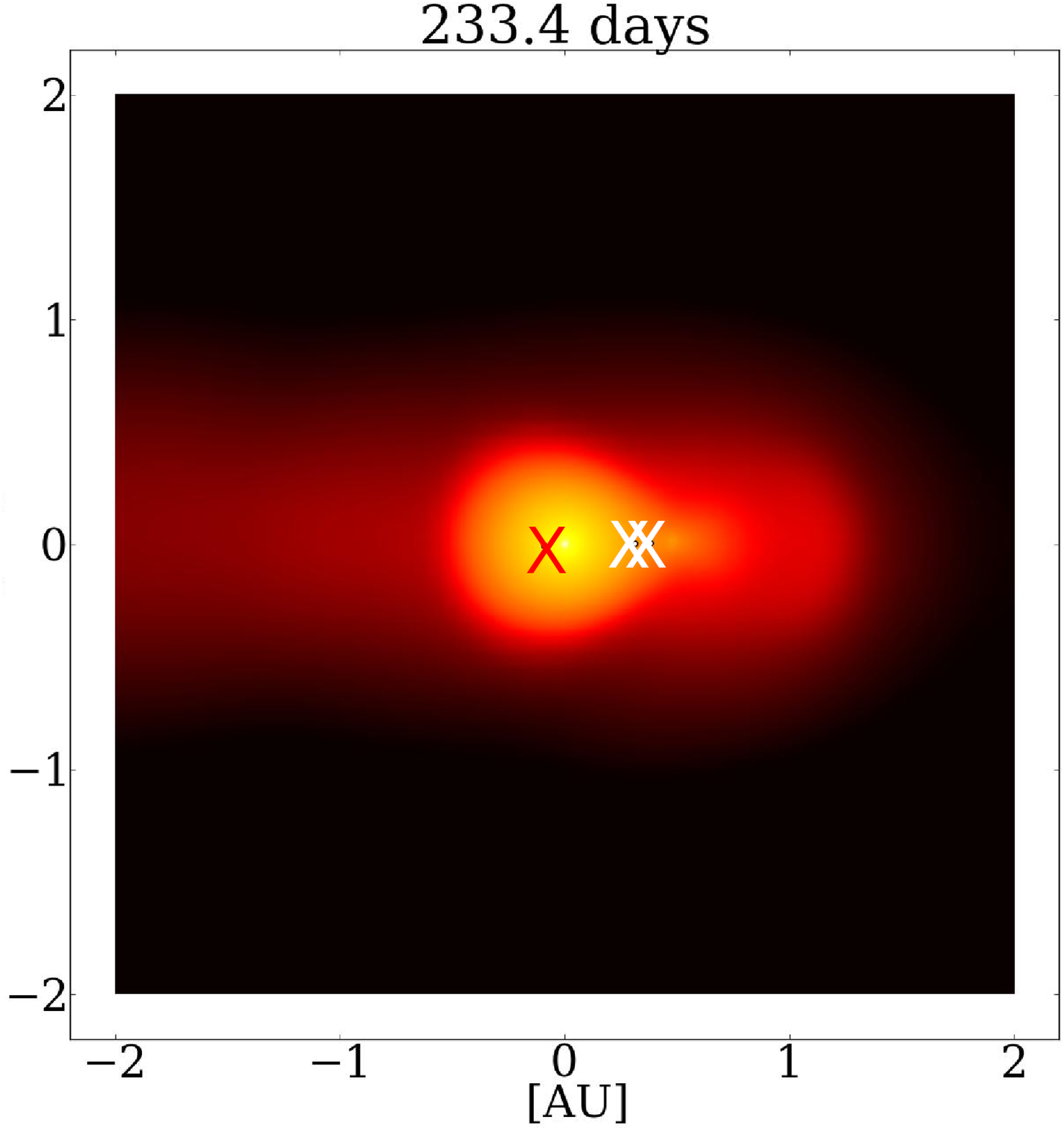}\includegraphics[width=0.09\columnwidth]{80265/colorbar}}\tabularnewline
	\multicolumn{3}{|c|}{\includegraphics[width=0.415\columnwidth]{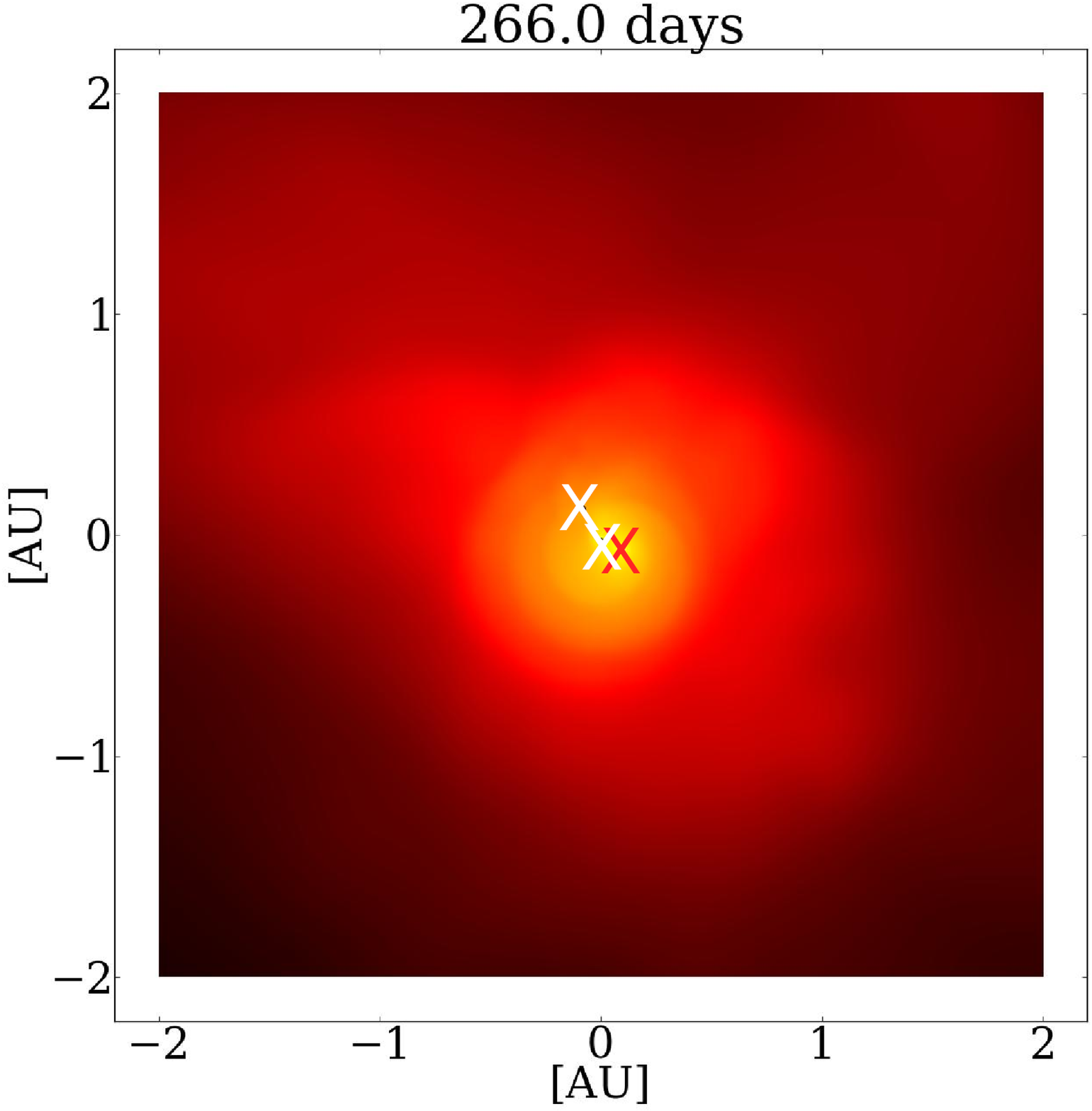}\includegraphics[width=0.4\columnwidth]{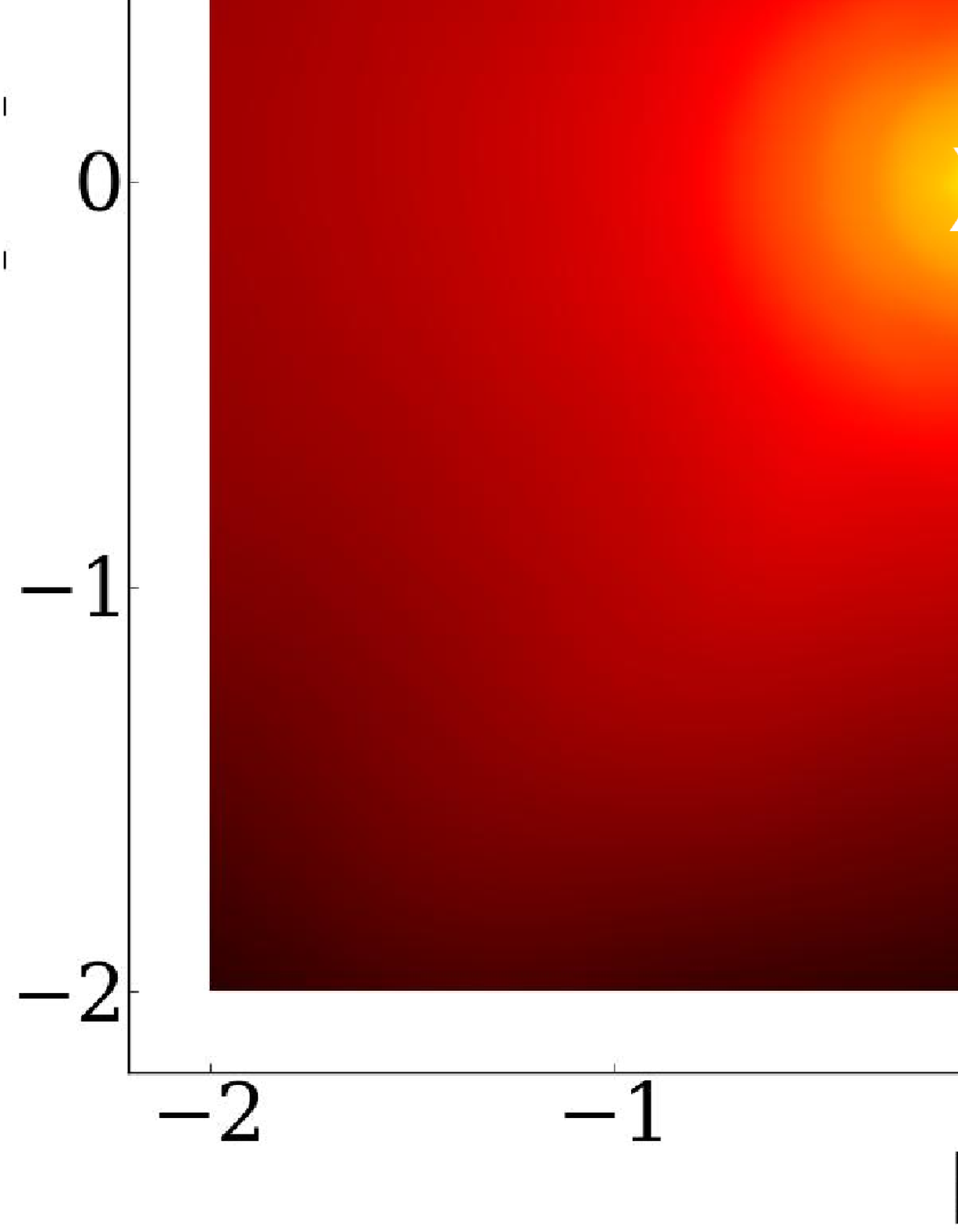}\includegraphics[width=0.09\columnwidth]{80265/colorbar}}
	\end{tabular}
	\caption{\label{tab:snapshots} Simulation 1 in Table \ref{tab:system-configurations} - a common envelope evolution of an $8M_{\odot}$
		giant with an envelope radius of $0.5\text{AU}$, orbited by an inner-binary ($\text{binary}_{2-3}$)
		composed of two $1M_{\odot}$ main- sequence stars initially positioned
		at $1\text{AU}$. The left and right panels correspond to a face-on
		view ($\text{binary}_{2-3}$ moves on an anti-clock-wise orbit) and an edge-on
		view (the binary moves towards us), respectively. The components of
		the components of $\text{binary}_{2-3}$ are marked with white 'X' symbols, and the giant
		core is marked by a red 'X' symbol. The symbol sizes do not correspond
		to the stellar sizes and are just shown for clarity. }
\end{figure}

Initially, $\text{binary}_{2-3}$ is located outside the stellar envelope
of the red-giant. It then progressively in-spirals due to the 
interaction with the stellar envelope. Figure \ref{tab:snapshots}
shows different snapshots of the density profile during the evolution
of the CE of the first simulated scenario (see Table \ref{tab:system-configurations}). When the binary spirals into the envelope and
forms a TCE, the spiral-in becomes more rapid, the envelope expands,
and mass-loss ensues mostly through the second Lagrangian point. If
we compare the last snapshots with those of a corresponding (same mass) binary
system (simulations 1-10 and 14 in Table \ref{tab:system-configurations}), the shape of the surrounding gas differs significantly between
the two cases (see bottom of Figure \ref{fig:binaryVStriple}). The TCE case is far less symmetric than the binary
CE case (see also \citealt{1992AJ....104.2151S,2017ApJ...837L..10B, 2019MNRAS.490.4748S}
in this regard). 

\begin{figure}
\includegraphics[width=\linewidth,clip]{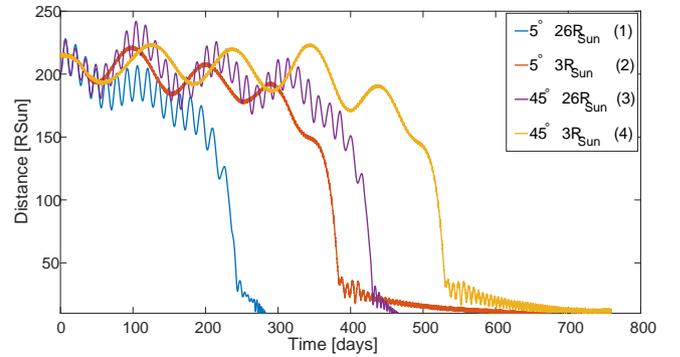}
\par
\caption{\label{Distances0Phase}Evolution of the distance between the
center of mass of $\text{binary}_{2-3}$ and the giant core, for simulations 1-4 (see Table \ref{tab:system-configurations}). The plunge-in phase begins earlier in low inclination
cases, due to the effective closer initial distance of one of the
companions to the giant core. A shorter $\text{binary}_{2-3}$ separation extends the
duration of the self regulated phase, because of the potential energy
stored in the orbit of $\text{binary}_{2-3}$ }, part of which is extracted by the gaseous envelope
during the in-spiral, leading to further envelope extension and mass
ejection, and thereby slowing the in-spiral onto the giant core. 
\end{figure}

\begin{figure}
\raggedright{}\includegraphics[width=\linewidth,clip]{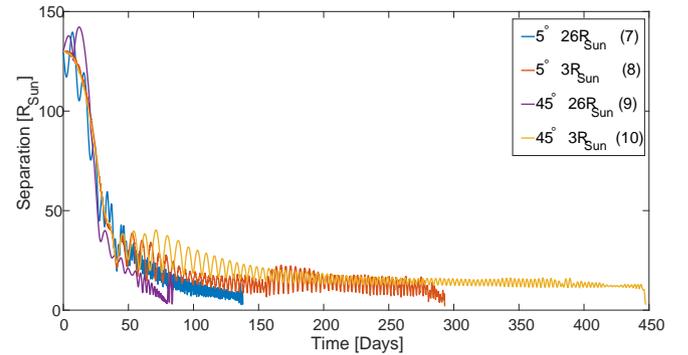}\caption{\label{fig:separation90}Evolution of the distance between 
the COM of $\text{binary}_{2-3}$ and the core for systems 7-10 in Table \ref{tab:system-configurations}}, with initial $90^{\circ}$
phase-angle (the components of $\text{binary}_{2-3}$ orbits their center of mass on a perpendicular plane than their mutual motion around the giant. In addition, due to the orbital phase, both companions are initially
at the same distance from the giant's core). 
\end{figure}

When comparing cases of inner-binaries ($\text{binary}_{2-3}$) with initially shorter and
long periods we find that the more compact binaries we considered
in-spiral more slowly than the corresponding models with longer periods
(see Figures \ref{Distances0Phase} and \ref{fig:separation90}).
The former systems are more strongly bound, and are not disrupted as they
in-spiral close to the giant core; rather we find that in such cases
the components of $\text{binary}_{2-3}$ in-spiral and merge with each other before
approaching the core. Conversely, the latter, wider $\text{binary}_{2-3}$
in-spiral into the giant core more rapidly, and are disrupted due
to the interaction with the central potential, leading to a chaotic
triple dynamics of the two (former) $\text{binary}_{2-3}$ components and the
giant's core. The chaotic evolution can then lead to the merger of any
two of the components, and the possible ejection of one of them from
the system.

In Figure \ref{Distances0Phase} we compare the separations between
the center of mass of $\text{binary}_{2-3}$ and the giant's core for the
configurations initialized with $0^{\circ}$ orbital phase. All of
our simulations of such initial zero-phase cases resulted in a core
merger with one of the components of $\text{binary}_{2-3}$. This suggests that the evolution
is sensitively dependent on the initial closest approach of the binary
closer component. The phase, and hence the initial separation of the
closest binary component to the giant determine the evolution of the
in-spiral, its duration and timing of both the entrance to the rapid
plunge-in phase as well as the later stage of the CE,
before merger. For binaries initialized with $90^{\circ}$ inclination
in respect to the orbit (orbital phase), both the stellar components
are effectively initially at larger separations from the giant, and
therefore show far weaker interaction at first, and required the extension
of our simulation run times. Therefore, in such cases, we re-initialized these models and placed $\text{binary}_{2-3}$ closer to the edge of the envelope, at
$\sim0.6\text{AU}$. The results of this configuration can be seen
in Figure \ref{fig:separation90}, where, similar to the low-inclination
cases, the evolution of $\text{binary}_{2-3}$ with smaller inner separations lead
to their mutual merger before the binary approached the core. 

\begin{figure}
\includegraphics[width=\linewidth,clip]{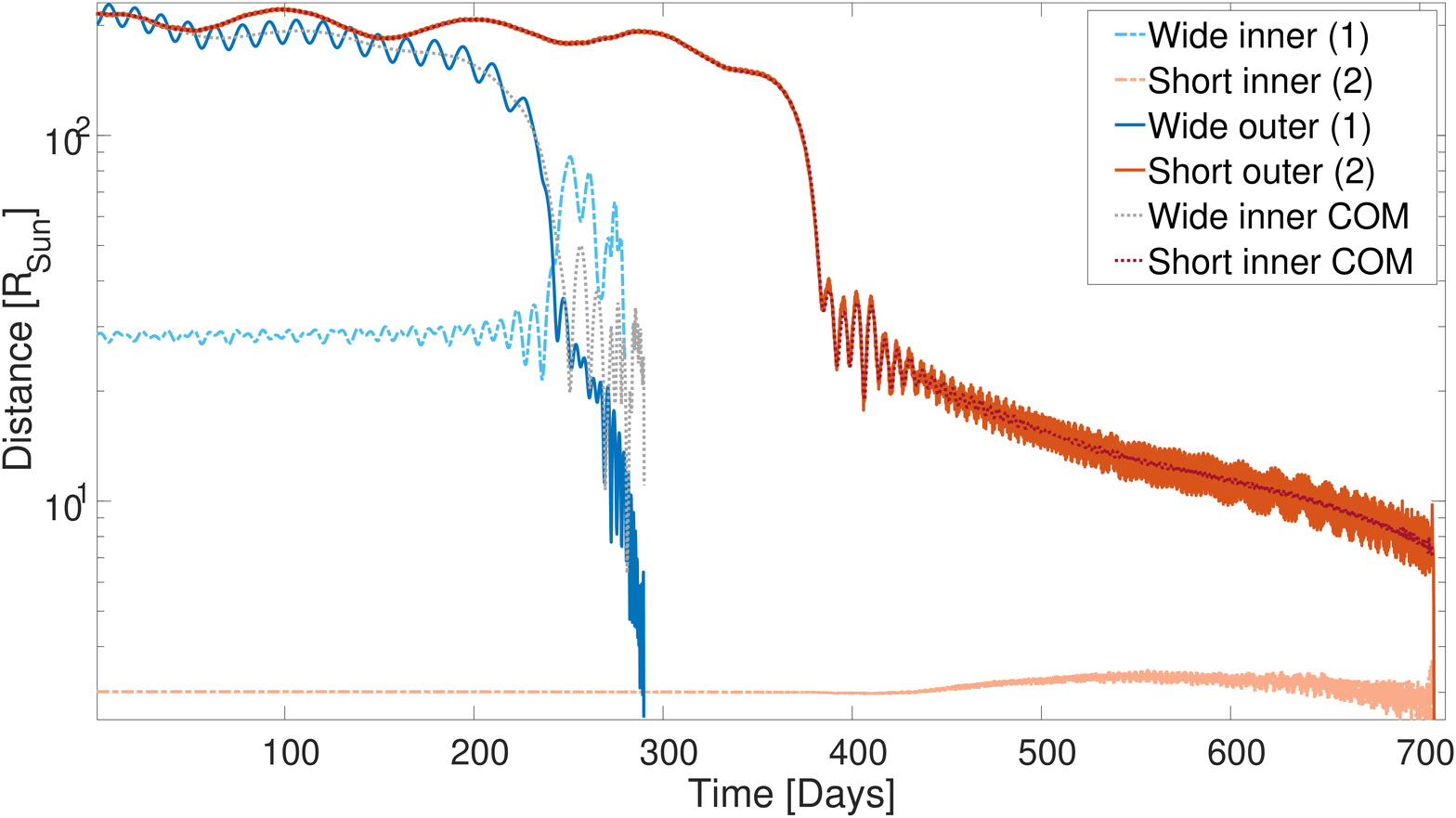}
\par
\includegraphics[width=\linewidth,clip]{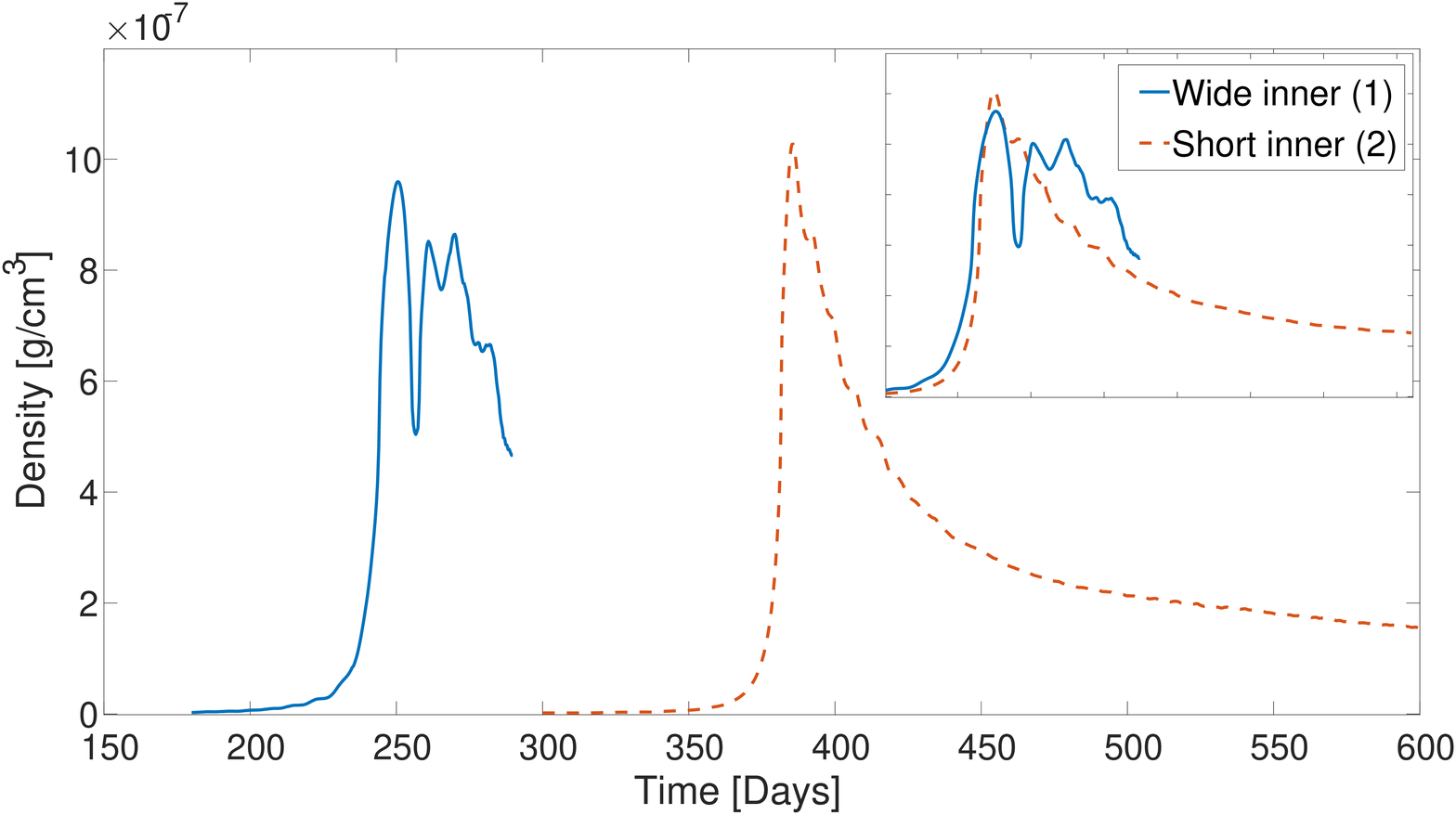}
\par
\includegraphics[width=\linewidth,clip]{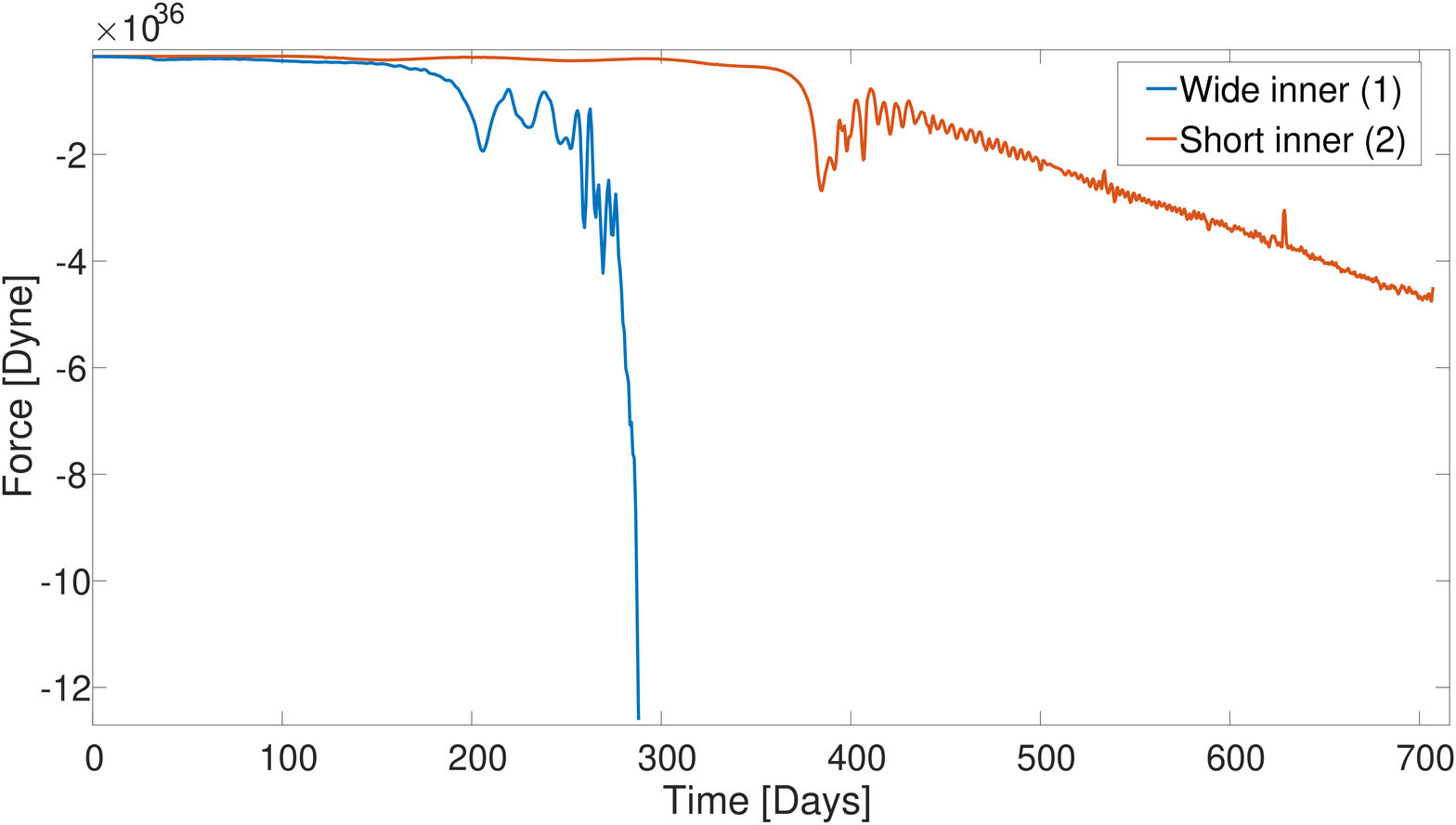}
\par
\caption{\label{InnerSepComp}  Evolution of properties of the triple components (for models 1 and 2). 
Upper panel: Evolution of the outer separation ($\text{binary}_{1-23}$, i,e- $\text{binary}_{2-3}$'s COM and core) and the inner separation of $\text{binary}_{2-3}$ (dashed lines) during the evolution. Both the separation between the COM of $\text{binary}_{2-3}$ (dotted lines) and the separation between its closest component to the core (solid lines) are shown. The wider inner-binary ($\text{binary}_{2-3}$) is disrupted as it reaches the region of the core, whereas the $\text{binary}_{2-3}$ with the shorter-period (orange) remains bound, imparts more energy to the envelope expanding it and decreasing its density, until eventually it merges. 
Middle panel: The local density around the center of mass of $\text{binary}_{2-3}$. We chose a sphere with a radius of $50 R_\odot$ to calculate the average density around $\text{binary}_{2-3}$'s COM. The average density around the more compact $\text{binary}_{2-3}$ decreases more rapidly, although there is an apparent strong decrease of the density around the wider binary close to its merging point. At this stage, the location of the COM is farther from the dense region near the giant core, due to the binary disruption (see upper panel). 
Lower panel: The total gravitational force on the $\text{binary}_{2-3}$. This force increases rapidly as the wider binary in-spirals to the core, while a slower evolution is seen for the more compact binary, due to its stronger dilution of the envelope throughout the in-spiral. }
\end{figure}

For the two different inclinations we considered, the in-spiral process
lasted much longer for the more compact $\text{binary}_{2-3}$ (See Figure \ref{InnerSepComp}). This is due to additional energy imparted to the envelope by $\text{binary}_{2-3}$
as its two components mutually in-spiral through their coupling to the
gaseous envelope (on top of the in-spiral of the center of mass of $\text{binary}_{2-3}$
onto the RG core). In other words, the in-spiral of $\text{binary}_{2-3}$
provides an additional energy/momentum source term. The expansion
of the CE is accelerated and its density decreases, consequently
decreasing the dissipation and in-spiral of $\text{binary}_{2-3}$'s COM onto
the RG core. In both inclinations, the shorter period $\text{binary}_{2-3}$ resulted
in an inner merger just shortly before merging with the core. In contrast,
binaries with larger separation were eventually disrupted as they
inspiraled close to the core.  

In Figure \ref{InnerSepComp} we compare the evolution of systems with different initial inner $\text{binary}_{2-3}$ separations (simulations 1 and 2 in Table \ref{tab:system-configurations}, one with a wider inner binary ($26 R_{\odot}$) and one with a short period inner binary ($3 R_{\odot}$). The separation of the outer binary of the wider inner binary decreases more rapidly, and its inner separation increases when reaching the central region of the giant, leading to its disruption. The larger binding energy of $\text{binary}_{2-3}$ with initial shorter orbital period gives rise to a slower decrease of its outer separation, a faster dilution of the envelope (middle panel), and a weaker gravitational force on $\text{binary}_{2-3}$ (lower panel). As a result, the components of $\text{binary}_{2-3}$ merge before reaching the core of the giant.

\begin{figure}
\includegraphics[width=\linewidth,clip]{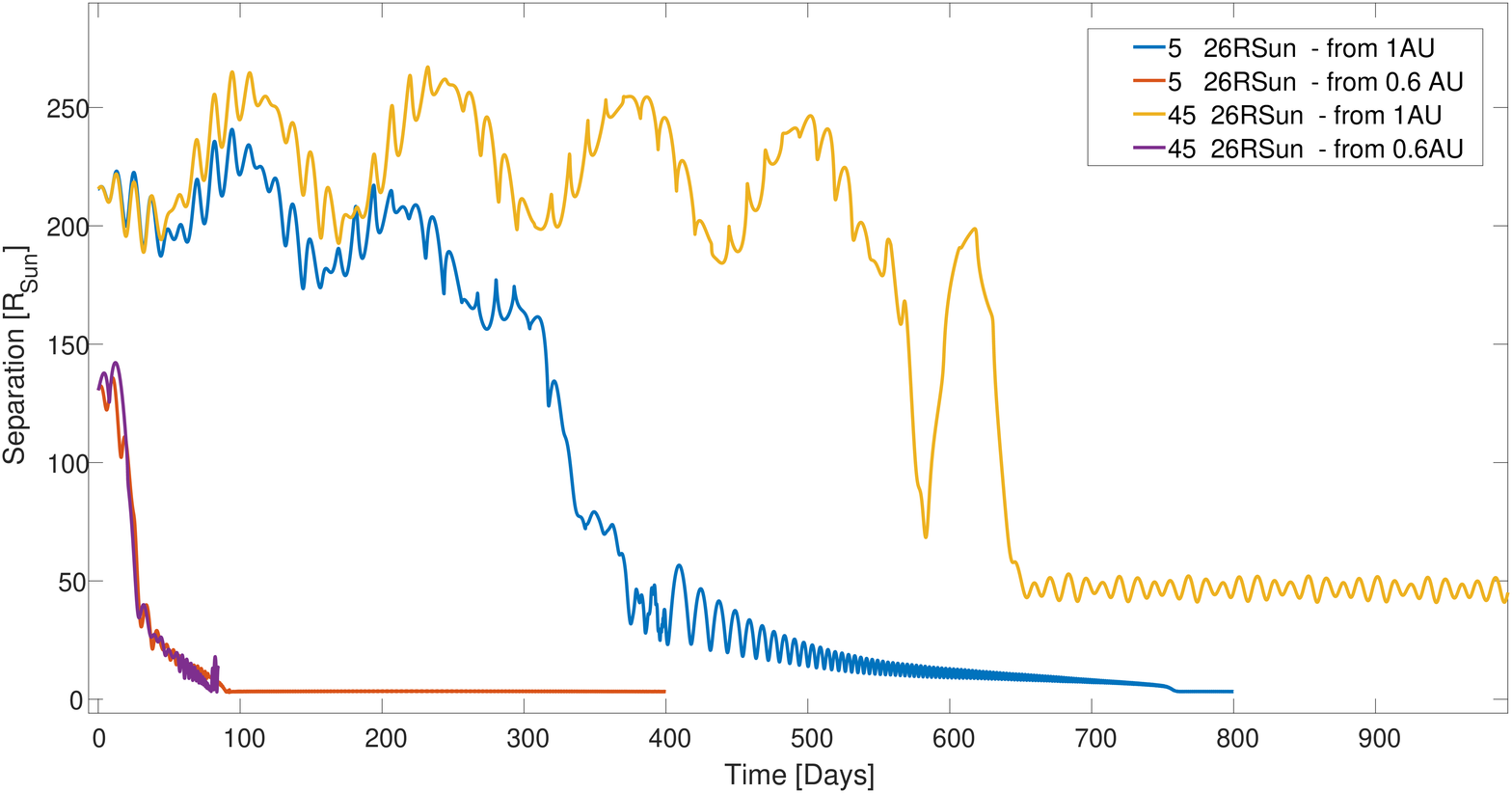}\caption{\label{fig:farAndClose}The separations between the $\text{binary}_{2-3}$ and
the core for systems with a $90^{\circ}$ phase, for both initial
positions - at 1AU (simulations 5,6), and 0.6AU (7,8 in Table \ref{tab:system-configurations}}
\end{figure}

In order to consider the sensitivity to the initial separation between
the $\text{binary}_{2-3}$ and the giant we studied the differences between the simulations
where the inner binaries ($\text{binary}_{2-3}$) where positioned at 1AU, and those initiated
at 0.6AU. Figure \ref{fig:farAndClose} presents the results. We find
that $\text{binary}_{2-3}$ significantly evolves in the 1 AU models before
they even reach separations of 0.6 AU, and can even merge before reaching
that point. In other words, it is critical to initialize the binaries
sufficiently far from the giant core as to correctly follow their
evolution, as significant evolution can happen even in the early in-spiral phases  (in agreement with results obtained for binary CE by \citealt{Iac+17,Reichardt19}).  

\begin{figure}
\centering
\includegraphics[width=0.7\columnwidth,clip]{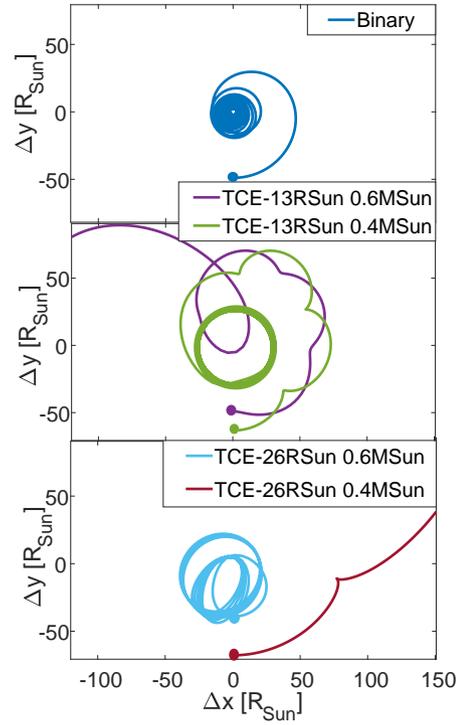}
\caption{\label{fig:OrbitsOf2M+1M} Comparison of the orbital evolution in
a common envelope for triples with mass components $2M_{\odot}+0.6M_{\odot}+0.4M_{\odot}$
of different inner $\text{binary}_{2-3}$ separations, and the CE evolution of the corresponding
binaries of equivalent masses (simulations 12,13 and 15 in Table \ref{tab:system-configurations}). The first companion is $0.6M_{\odot}$compact
object, whereas the second is $0.4M_{\odot}$. $\Delta x$ and $\Delta y$ are the difference in the location between the companion and the giant core in $x$ and $y$ coordinates.The dots mark the initial separation of the companion from the core.}
\end{figure}
\begin{table}
\begin{tabular}{|c|c|c|c|c|}
\hline 
Sim. & Inclination & $a_{\text{in}}$  & Merged & Merger time 
\\ &&($R_{\odot})$&components&(days)\tabularnewline
\hline 
\hline 
1&$5^{\circ}$ & $26$ & companion & 289\tabularnewline
&&& + core&\tabularnewline
\hline 
2&$5^{\circ}$ & $3$ & $\text{binary}_{2-3}$ & 706\tabularnewline
\hline 
3&$45^{\circ}$ & $26$ & companion & 475\tabularnewline
&&& + core&\tabularnewline
\hline 
4&$45^{\circ}$ & $3$ & $\text{binary}_{2-3}$ & 760\tabularnewline
\hline 
14&\multicolumn{2}{|c|}{Binary} & core merger & 106\tabularnewline
\hline 
\end{tabular}

\caption{\label{tab:Results-summary-of820}Results summary of $8M_{\odot}+1M_{\odot}+1M_{\odot}$
with $0^{\circ}$ phase All models here are initialized with the $\text{binary}_{2-3}$ located at a distance of 1AU from the giant core.}
\end{table}
\begin{table}
\begin{tabular}{|c|c|c|c|c|}
\hline 
Sim. & Inclination & $a_{\text{in}}$ & Merged & Merger time 
\\ &&($R_{\odot})$&components&(days)\tabularnewline
\hline 
\hline 
7 & $5^{\circ}$ & $26$ & companion & 544\tabularnewline
&&& + core&\tabularnewline
\hline 
8 & $5^{\circ}$ & $3$ & $\text{binary}_{2-3}$ & 75\tabularnewline
\hline 
9 & $45^{\circ}$ & $26$ & companion & 85\tabularnewline
&&& + core&\tabularnewline
\hline 
10& $45^{\circ}$ & $3$ & $\text{binary}_{2-3}$ & 5\tabularnewline
\hline 
\end{tabular}
\caption{\label{tab:Results-summary-of8290}Results summary of $8M_{\odot}+1M_{\odot}+1M_{\odot}$
with $90^{\circ}$ phase. All models here are initialized with $\text{binary}_{2-3}$ located at a distance of 0.6AU from the giant core.}
\end{table}

\begin{table}
\begin{tabular}{|c|c|c|c|}
\hline 
Sim. & $a_{\text{in}}$ & Ejected component & Ejection velocity\\ 
&($R_{\odot})$&&(km s$^{-1}$)\tabularnewline
\hline 
\hline 
12 & $13$ & $0.6M_{\odot}$ & $\sim95$\tabularnewline
\hline 
13 & $26$ & $0.4M_{\odot}$ & $\sim120$\tabularnewline
\hline 
\end{tabular}

\caption{\label{tab:Results-summary-of210} Results summary of the models with $2M_{\odot}+0.6M_{\odot}+0.4M_{\odot}$ components,
all with $0^{\circ}$ inclination and orbital phase. In both separations considered, no merger between any of the components occurs during the simulation.
In the case of the small separation of $3R_{\odot}$ $\text{binary}_{2-3}$
merged very rapidly. Ejection velocities were measured upon leaving the envelope. }
\end{table}

\subsection{Binaries vs. triples}
\begin{figure}
\centering
\includegraphics[width=\linewidth,clip]{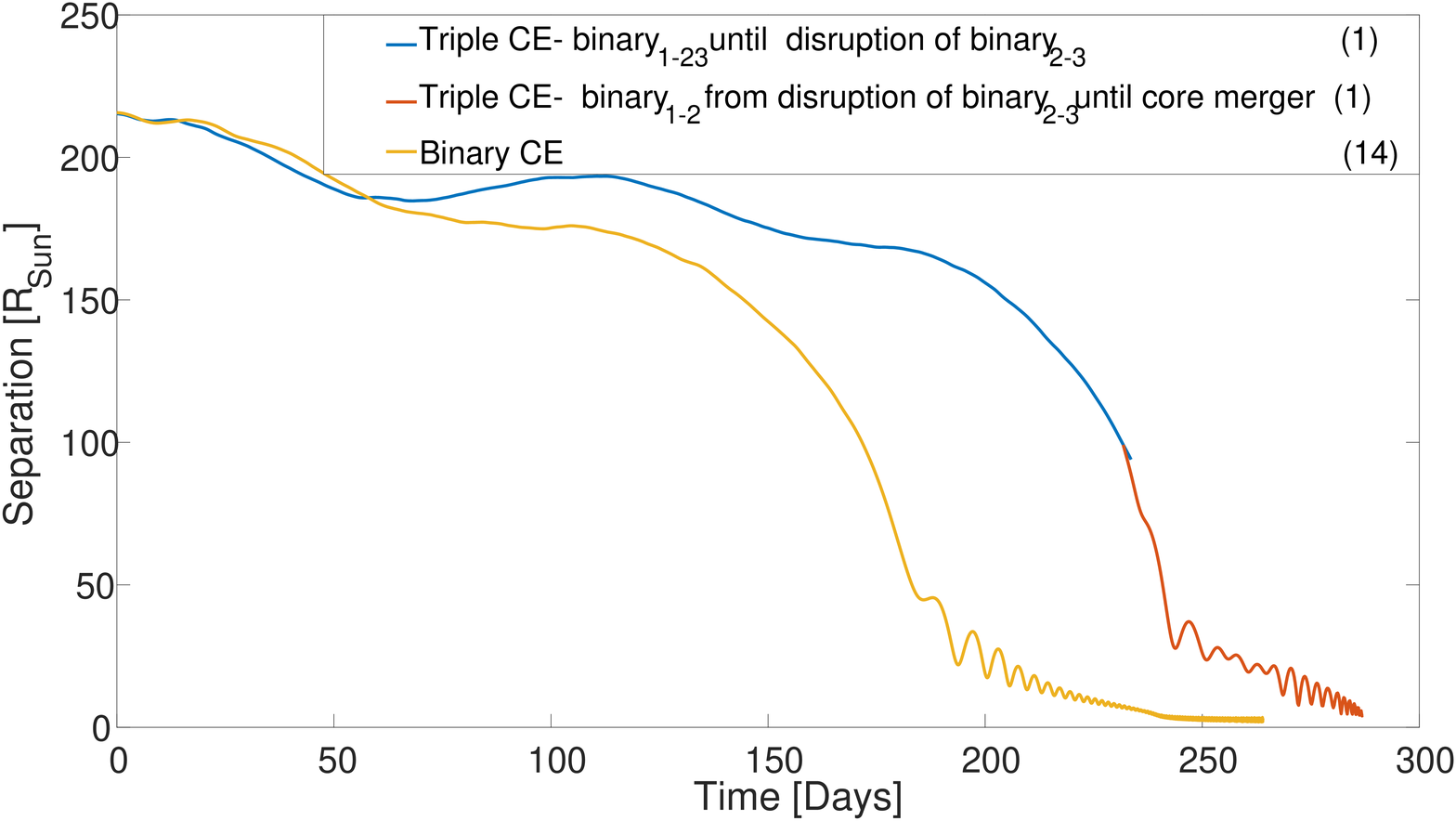}
\par
\includegraphics[width=\linewidth,clip]{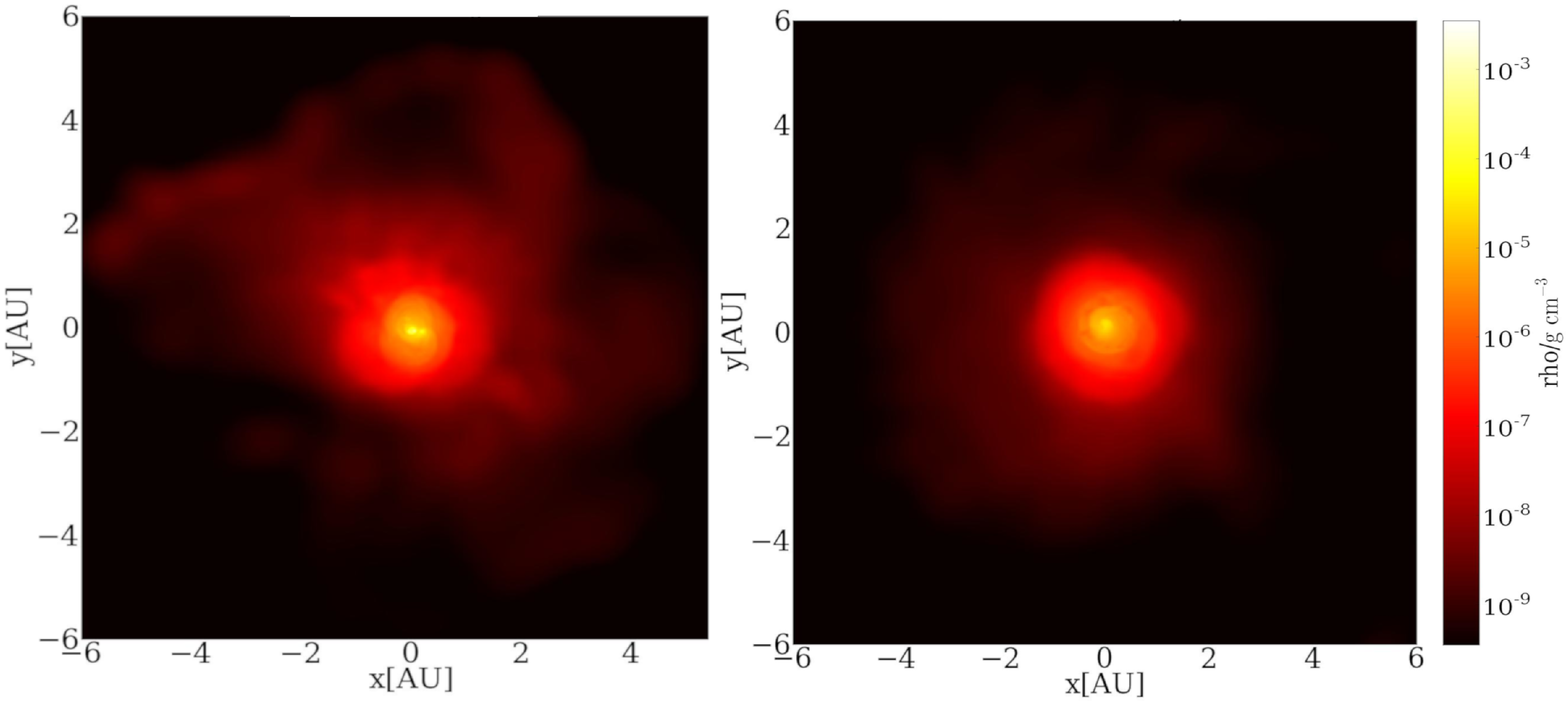}
\par
\includegraphics[width=\linewidth,clip]{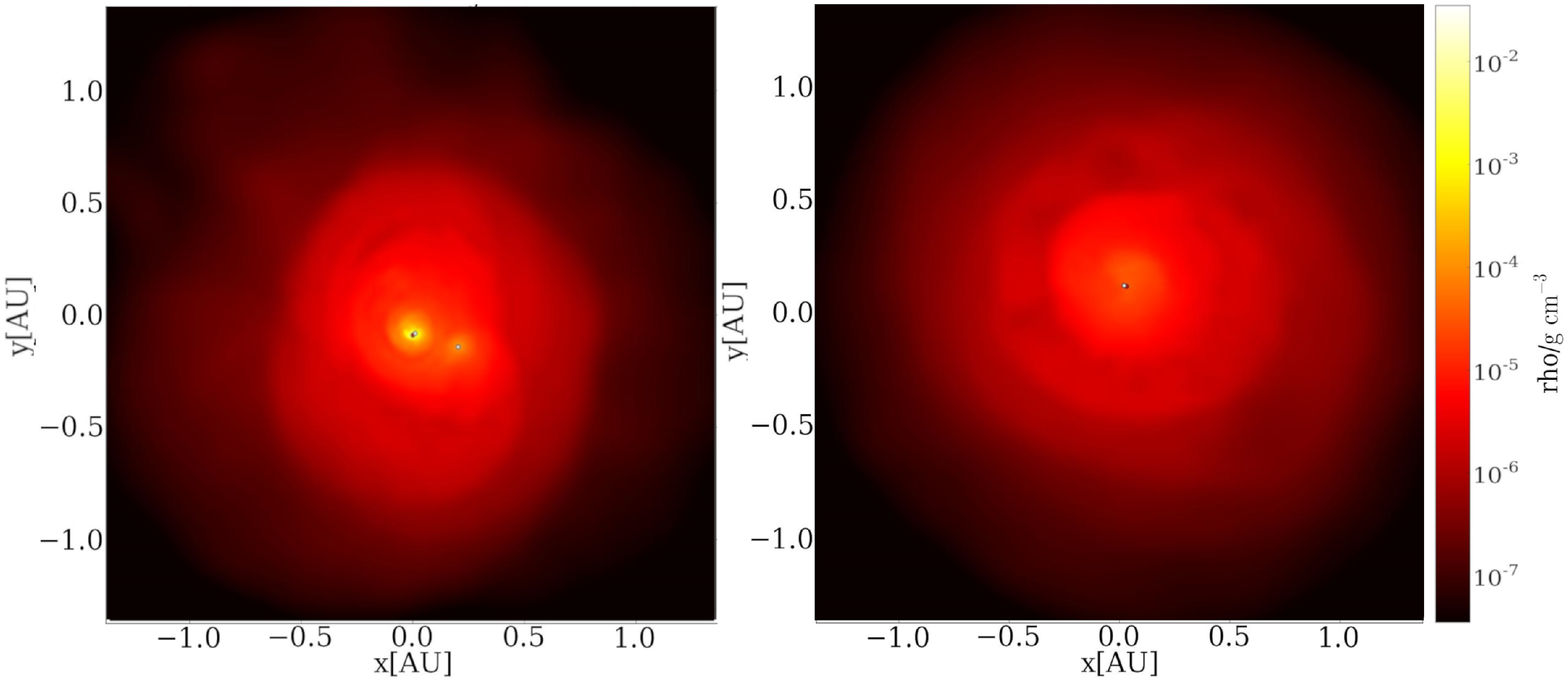}
\par
\caption{\label{fig:binaryVStriple}Comparison between the evolution of a
triple system with $8M_{\odot}$ giant with $1M_{\odot}+1M_{\odot}$ companions (simulation 1 in Table \ref{tab:system-configurations})
and a corresponding binary system, consisting of a companion which is
the sum of both components of $\text{binary}_{2-3}$ (simulation 14).
Upper plot: the separation between the COM of $\text{binary}_{2-3}$ and the core, until the disruption of $\text{binary}_{2-3}$'s components, follows by the separation between the companion to be merged with the core and the giant's core.
Middle: snapshots that were taken at the end of simulations 1 and 14 in Table \ref{tab:system-configurations} from left to right, to demonstrate the different shape of the remnants.Bottom: same snapshots, but zoomed and different density scale. We can see that even the denser part of the envelope is much less symmetric in the left snapshot of the TCE, than the one on the right, of the parallel binary CE.}
\end{figure}

The general differences between TCEs and binary CEs can be studied
by comparing simulations of triple systems and their corresponding
binary systems, in which $\text{binary}_{2-3}$ is replaced by single star
having the summed mass of the binary components, initially positioned
at the COM of the original binary. Figure \ref{fig:binaryVStriple}
shows the comparison of the evolution of the separation between the
giant core and the companion star/binary-COM for simulations 1 and 14. As can be seen, the
TCE evolution extends longer than the binary CE evolution, although
it appears that the evolution following the fast plunge-in is not
significantly different. Similar conclusions can be obtained from
Figure \ref{fig:OrbitsOf2M+1M}. \\
The fraction of unbound mass in our triple simulations, which is still much less than the entire envelope, was between 11 and 28 percent of the envelope mass, slightly larger than the unbound mass found in our binary simulation (number 14) which was around 10 percent. 

\par

\section{Discussion}
\label{sec:Discussion}
Our results suggest TCE evolution in circumstellar configurations
typically lead to either the merger of the components of $\text{binary}_{2-3}$ before it
approaches the core; or the excitation of the orbit of $\text{binary}_{2-3}$  and
eventually its disruption as it in-spirals closer to the core, leading
to chaotic triple dynamics involving all three components (the components  of $\text{binary}_{2-3}$ and the giant core), which are still embedded in the gaseous
envelope. Following the chaotic evolution, the components of the disrupted
binary can either mutually merge, one or both can merge with the core,
or one of them may be ejected from the system. We briefly discuss
each of these possibilities below. 

As discussed above, the specific evolution of the TCE strongly depends
on the particular triple configuration adopted, including the inner and outer binary
separations, the mutual inclination between $\text{binary}_{2-3}$ and its
orbit around the giant ($\text{binary}_{1-23}$), and the masses of the components. 
We see that all of the parameters we investigated affect the triple
common envelope process; its duration, the final result and even its final ejecta's
observed shape. The results may have important implications for the
formation and evolution of various types of compact binaries, their
mergers and the possible electromagnetic and gravitational-wave transients
they might produce. 

\subsection{Mergers}
In our simulations, the exact nature of the components of $\text{binary}_{2-3}$
was not assumed, and they could potentially be either MS stars
or compact objects. A compact $\text{binary}_{2-3}$ that is consisted of 2 MS stars, overfill their Roche-lobe while the orbit becomes tighter. In this case, modeling $\text{binary}_{2-3}$ as 2 point-mass stars is not accurate to explore their fate accurately. However, if one is only interested whether they merge or not, this description is sufficient. We briefly discuss some of the more unique outcomes
that may potentially arise from such evolutionary scenarios. 

The merger of two MS stars could leave behind a blue straggler. Formation
of blue-stragglers in triples were explored by us and others before \citep{2009ApJ...697.1048P,per+12c,Nao+14},
but following very different evolutionary scenarios, and giving rise
to different outcomes. In particular, if the merged star does not
merge with the core during the CE, the post-CE binary (formerly triple)
would become a unique binary - a potentially short-period blue-straggler
binary, with a likely He-WD companion (or He-CO hybrid WD; \citealp{2019MNRAS.482.1135Z}),
a configuration which is difficult to explain through other evolutionary
scenarios. Interestingly, a binary He/hybrid-WD - blue-stragglers might have
already been observed \citep{wd-BS-detection}.

If the components of $\text{binary}_{2-3}$ are two white-dwarfs, the merger may
leave behind a massive WD, that may later merge with the RG core during
the CE, or survive and then potentially merge with the remnant of
the RG core, likely a He-WD or hybrid He-CO WD (if they in-spiral
through gravitational-wave emission). Such evolution might give rise
to a type Ia supernova \citep{2019arXiv191007532P}. Alternatively,
the merger of $\text{binary}_{2-3}$ WD might result in type Ia supernova
- (see also \citealp{DiStefano2019}) or form a different type of
star (e.g. \citealp{stefano2018mass}). In such cases, the supernova
would occur while still embedded in the CE. The strong shock interaction
with the envelope might produce a long-lasting and more luminous supernova,
possibly also related to the recently suggested origin of super-luminous
supernovae from thermonuclear explosions inside a common envelope
\citep{2020Sci...367..415J}.

We note in passing, that in cases where $\text{binary}_{2-3}$ is composed of neutron stars or black
holes, a TCE could induce their merger, leading to the production of
gravitational-waves sources with unique signatures (e.g. somewhat
similar to the cases of CE-induced gravitational-waves sources explored
by us; \citet{ginat2019gravitational}). However, the evolution of
such massive components is not explored by our current models, and
the study of whether a realistic evolutionary scenario can produce
such cases is beyond the scope of the current work.

The result of a merger between one of the components of $\text{binary}_{2-3}$ and the RG core, leaves behind the second component, which can then
continue to a second CE phase. Such evolution will form a new star
with a larger mass than the original core, but smaller than the initial evolved
giant. The exact nature of such rejuvenated red-giant (or possibly
a Thorne-Zitkow, \citealt{1977ApJ...212..832T}; in case a neutron-star
in-spirals to the core) is yet to be explored. 

We should also note that in a somewhat different scenario of a resulting
binary system, a further accretion could occur from the inner gas
with its new formed core, on the other companion, suggested to form
an X-Ray binary by \citet{1986MNRAS.220P..13E}. 
Further outcomes of the merger of $binary_{2-3}$ components have also been suggested and studied by   \citet{2016MNRAS.455.1584S,Hillel2017} and \citet{NoamSNReview2019}.

\subsection{Ejections, runaway stars and single SdB stars}
Due to the chaotic triple interaction between the components of $\text{binary}_{2-3}$ and the RG core, one of the components might be ejected. Its typical
velocity would be comparable to the orbital velocities at the point
of the disruption of $\text{binary}_{2-3}$, which can be as high as a few tens
or even 100 km s$^{-1}$. The TCE could therefore give rise to a novel
channel for the production of runaway stars, albeit likely only in
relative rare cases. In simulation 12 (see Table \ref{tab:system-configurations} for its configuration, Figure \ref{fig:OrbitsOf2M+1M} for its orbit and Table \ref{tab:Results-summary-of210} for the summary of results), the more massive companion was ejected with an initial (relative to the core) velocity of $v_{dist}\approx133\text{km}\cdot \text{s}^{-1}$, and then went through the giant envelope, dissipating some of the velocity to reach a velocity of $v_{ej}\approx95\text{km}\cdot \text{s}^{-1}$ upon leaving the envelope. A simple calculation of its velocity without the effect of dynamical friction would give  a velocity of about $\sim129\text{km}\cdot \text{s}^{-1}$, higher than the measured velocity, showing the importance of the CE dissipation even during ejections. Ejections are therefore less likely, or give rise to lower ejection velocities when interacting with more massive envelopes. If the RG core is ejected (i.e the central core is ejected while the envelope material remains bound to the $\text{binary}_{2-3}$ or to one of its components), it might be observed
as a single sdB star. Interestingly, single sdB stars are difficult to explain as such stars
are typically expected (and observed; e.g. \citealt{2008ASPC..392..207G})
to have a close-by companion which took part in their formation through
stripping their envelope. Though TCEs are unlikely to explain a high frequency
of single sdBs, the finding of runway single sdBs could provide
a potential smoking gun signatures for such processes.

\subsection{Planetary nebulae}
Shortly after the end of the self regulating phase, any of the observed
systems will consist of one or more compact objects, surrounded by
the unbound gas as a planetary nebula. Planetary nebula could be the result of a post-AGB star that lost its envelope during late-evolution stages. However, the ejection of an AGB stellar envelope is likely to be spherically symmetric, thereby producing a symmetric planetary nebula, whereas the mass loss in common envelope evolution is mostly via the second and third Lagrangian points, both during the CE as well as during the interaction prior to this phase. The resulting planetary nebulae of such binary system is therefore not expected to be spherically symmetric, and could even sometimes be bipolar \citep{2008demarcoPN,2009dematcoshapingpn,2009binaryPNlowIon,2014jonespn,jones15-postCEbinaryinsidenebulae}. Moreover, \cite{2020MNRASZouFrankAsymmetricPN} showed that even spherical outflows from the post-CE binary can be highly deflected by the interaction with the CE ejecta, and result in highly collimated bipolar outflows that may increase the asymmetry of the planetary nebula.
As discussed above, TCE could
give rise to highly aspherical and non axis-symmetric planetary nebulae, and the shape might not be ellipsoidal as in the post-CE binary cases, thereby giving rise to a wide diversity of shapes \citealt{1992AJ....104.2151S,2016MNRAS.455.1584S,2017ApJ...837L..10B}). 

\subsection{Mass loss}
Hydrodynamical simulations of binary CEE show that only a fraction
of the envelope mass is ejected, while the majority (typically 90-80$\%$)
remains bound; (e.g. \citealt{Pass+12,Ric+12,Iva+13,Iva+15,Kur+16,Ohl+16,Iac+17}),
posing a potential problem, since these should result in a merger of the entire system instead of a compact post-CE binary. It was suggested that recombination energy can provide an
additional energy source to drive the ejection of the envelope \citep[and references therein]{Iva+15}.
However, the fraction of the recombination energy lost to radiation
is still debated \citep{Sok+03,2011ASPC..447...91I,Cla+17,2017MNRAS.472.4361S,2018MNRAS.478.1818G}.
Furthermore, the existence of wide post-CE orbits cannot be explained by merely including recombination energy \citep{Reichardt2020}. Others suggested that accretion energy mediated by
jet/outflows may play a role \citep[e.g.][and references therein]{2019MNRAS.488.5615S,2019MNRAS.490.4748S}
or that dust formation inside the CE could drive winds and help to
eject more material (\citealt{DustDrivenWindsPaper}, and references
therein) on longer timescales, with possible observational evidence
for such long-term mass-loss \citep{Mic+19,2019arXiv190710068I}.
Our hydrodynamical simulations of a TCE evolution show that TCE also
gives rise to inefficient mass-loss. However, as discussed above,
the coupling of the binding energy of $\text{binary}_{2-3}$ to the envelope
provides and additional energy/momentum source and leads to a longer
in-spiral timescale, and a much larger mass-loss from the TCE, compared
with the corresponding binary CEE cases. The unbound mass was calculated by the the total mass of the SPH particles with $e_{kin} + e_{pot} > 0 $, where $e_{kin}$ and $e_{pot}$  are the specific kinetic energy of the particle, with a velocity relative to the COM of the envelope, and an approximation of its gravitational potential due to the envelope. The gravitational potential energy between any two particles is calculated as follows: \[E_{pot,ij} = -Gm_im_j/r_{ij}\] where $m_i$, $m_j$ are the masses of particles i and j, and $r_{ij}$ is their separation. The potential energy of a particle is the sum over all particles potential in the system- \[E_{pot,i} = \sum_j E_{pot,ij}.\] Other works suggest the inclusion of the thermal energy of a particle in considerations of the ejected mass; however, this could only give an upper limit, as much of the thermal energy could be radiated away. Our models do not include radiative transfer and we do not consider the thermal energy in the mass-loss estimate. In any case, in this work, we do not try to solve the general problem of CE mass loss, but only compare between the different scenarios. Therefore, this approximation is applicable for our purposes.  In the longest-lasting in-spiral
we find a mass loss of $\sim27\%$, compared with only $\sim8\%$
in the equivalent binary case. Moreover, our simulations terminate
once two of the components merges (reach the sum of their radii), while the CE may proceed afterwards,
and therefore the TCE mass loss fractions cited are only a lower-limit.
Since only a fraction of CE cases involve triples, the more efficient
TCE mass-losses cannot generally solve the envelope-ejection problem,
but the more significant mass-loss do show an additional qualitative
difference in the TCE evolution compared to binary CEE. 

\section{Summary}
In this study we have carried out the first hydrodynamical modeling of a
triple common envelope evolution in a circumstellar configuration,
where a more compact binary (termed $\text{binary}_{2-3}$ or the inner binary) orbits an evolved
giant and eventually in-spirals into its envelope producing a TCE.
We made use of the Gadget2 SPH code coupled to few-body codes using
the AMUSE environment to combine the hydrodynamical aspects with the
few-body dynamics involved. Given the computational expense we studied
only a limited grid of models, serving as initial exploration of the
sensitivity of the evolution to the initial orbital configurations,
and the possible different outcomes of TCEs. We studied a total of
11 TCE models with different masses, inner-$\text{binary}_{2-3}$ separations, orbital,
relative inclinations and orbital phases. We also compared our models
with corresponding binaries, where $\text{binary}_{2-3}$ was replaced with
a single component of the same total binary mass. We terminated the
simulations once any two components merged during the simulation (the components of $\text{binary}_{2-3}$ and/or the RG-core).

We find that the TCE evolution leads to both the mutual in-spiral
of the components of $\text{binary}_{2-3}$, and their possible merger, as well
as the in-spiral of $\text{binary}_{2-3}$ towards the red-giant core. We
find that the more compact $\text{binary}_{2-3}$ configurations result in the
mutual merger of the components of $\text{binary}_{2-3}$ before they approach
the RG-core, while wider inner-$\text{binaries}_{2-3}$ do not merge, but in-spiral
to the core and are then disrupted by the RG inner potential of $\text{binary}_{2-3}$. In the
latter case the (now unbound) components of $\text{binary}_{2-3}$ and the RG core
evolve through chaotic triple dynamics, while still embedded in
the envelope, leading to the merger of at least two of these components,
and the possible ejection of the third. 
\\
The $\text{binary}_{2-3}$ provides an additional energy/momentum source, and
its coupling to the envelope gives rise to stronger expansion of the
envelope and significantly larger mass loss, but not substantial as to completely unbind the entire envelope. Consequently, the envelope
density decreases more rapidly, and the timescale for the in-spiral
towards the core is extended in comparison to the binary models. In
addition, this evolution lives behind a significantly more aspherical and non axis symmetric remnant than the binary case (see bottom of Figure \ref{fig:binaryVStriple}. We find that the specific evolution
is sensitive to the initial configuration, but our models provide
only a limited sample of the large phase space of triples, while a
full characterization of the dependence is yet to be explored.

Our findings suggest that TCE can give rise to unique outcomes, and
the possible production of peculiar blue-straggler binaries; unique
gravitational-wave sources with gas-coupling dominated evolution (see
also \citealt{ginat2019gravitational}); potentially super-luminous
peculiar thermonuclear supernovae (due to explosions following WD
mergers inside the TCE); short-GRBs from neutron-stars mergers inside
a TCE and the production of gravitational-wave sources; runaway stars
(and possibly runaway SdBs); and other exotic mergers and their potential
transient outcomes. Predicting the rates and branching ratios for
the rich phase space of TCEs is beyond the scope of our exploratory
study; and should be explored in the future.

\section*{Acknowledgements}
We greatly thank the referee, Orsola De Marco, for comprehensive and constructive comments. We thank Noam Soker for helpful discussions.
HG and HBP acknowledge support for this project from the European Union's Horizon 2020 research and innovation program under grant agreement No 865932-ERC-SNeX.
\newline
We used the following codes in the simulations, analysis and visualizations presented in this paper: AMUSE \citep{2009NewA...14..369P}, MESA (version 2208) \citep{2011ApJS..192....3P}, GADGET2 \citep{2005MNRAS.364.1105S}, HUAYNO \citep{2012NewA...17..711P}, MI6 \citep{10.1093/pasj/59.6.1095}, matplotlib \citep{matplotlib}, pynbody \citep{pynbody}, NumPy \citep{numPy}. All models were running on  the Astric computer cluster of the Israeli I-CORE center.

\section*{Data Availability}
All data underlying this research is available upon reasonable request to the corresponding authors.


\bsp	
\label{lastpage}
\end{document}